\documentclass[conference]{IEEEtran}
\IEEEoverridecommandlockouts
\usepackage{cite}
\usepackage{amsmath,amssymb,amsfonts}
\usepackage{algorithmic}
\usepackage[dvipdfmx]{graphicx}
\usepackage{textcomp}
\usepackage{amsmath,amssymb,enumerate,color,url}
\usepackage{bm}
\usepackage{caption}
\usepackage{subfigure}
\usepackage{subfigmat}
\usepackage{ascmac}
\usepackage{hyperref}
\usepackage{comment}
\usepackage{ulem}

\newcommand{\dd}{\mathrm{d}}
\newcommand{\e}{\mathrm{e}}
\newcommand{\ii}{\mathrm{i}}

\allowdisplaybreaks[1]

\def\BibTeX{{\rm B\kern-.05em{\sc i\kern-.025em b}\kern-.08em \kern-.1667em\lower.7ex\hbox{E}\kern-.125emX}}
    
\begin{document}
\title{Perturbative Expansion of the Fundamental Equation of Online User Dynamics for Describing Changes in Eigenfrequencies}

\author{\IEEEauthorblockN{Naoki Hirakura}
    \IEEEauthorblockA{
        \textit{Tokyo Metropolitan University} \\
Tokyo 191--0065, Japan \\
hirakura-naoki@ed.tmu.ac.jp}
\and
\IEEEauthorblockN{Masaki Aida}
    \IEEEauthorblockA{
        \textit{Tokyo Metropolitan University} \\
Tokyo 191--0065, Japan \\
aida@tmu.ac.jp}
}

\maketitle

\begin{abstract}
The oscillation model has been proposed as a theoretical framework for describing user dynamics in online social networks.
This model can model the user dynamics generated by a particular network structure and allow its causal relationships to be explicitly described.
In this paper, by applying perturbation theory to the fundamental equation of the oscillation model, we confirm that we can explicitly trace, at least in principle, the changes in user dynamics associated with changes in the network structure.
Specifically, we formulate perturbative expansions up to infinite order, by drawing on inferences from regularities found in perturbative expansions; the accuracy of perturbative expansions of finite order is evaluated by numerical experiments.
\end{abstract}


\section{Introduction}
In recent years, social media, such as Facebook and Twitter, are used daily for information gathering and communication.
While the widespread use of these services has contributed to improvements in the users' convenience, explosive user dynamics such as online flaming phenomena have become a frequent occurrence, and the negative impact on society cannot be ignored.
In particular, the one of most serious problems is that public opinion formed on the Internet has accelerated the division of real society. 
Therefore, understanding online user dynamics has become an important engineering issue. 
Unfortunately, actual user dynamics are considered too complicated to fully understand. 

There are two approaches to understand online user dynamics: data-driven and model-driven. 
The data-driven approach uses actual captured data to understand the specific case of the user dynamics observed.  
This paper focuses on the model-driven approach to identify the characteristics universal to various user dynamics. 
Our discussion is based on the framework of the oscillation model, which is a theoretical model for describing user dynamics in online social networks (OSNs)~\cite{MasakiAIDA20182017EBN0001}.
This model represents user dynamics as the wave equation on the network; it assumes that inter-user influence propagates at a finite speed in the network.
The wave equation is an equation that describes a phenomenon in which something propagates in a medium at a finite speed. 
In this case, it describes a situation in which a user's influence propagates to other users via OSNs.
The oscillation energy obtained from the solution of the wave equation represents the intensity of the network activity. Decomposing the overall energy into the oscillation energy of each node gives the generalized concept of node centrality, with the conventional node centrality (degree centrality and betweenness centrality)~\cite{freeman1978centrality,borgatti2009network} as a special case, which is well known in network analysis studies~\cite{TakanoIEICE.T.2018}.
Moreover, starting from the phenomenon of diverging oscillation energy, we can model explosive user dynamics such as online flaming phenomena and elucidate the conditions for their occurrence~\cite{MasakiAIDA20182017EBN0001,aidaASONAM2017}.
Of particular interest, if we tackle not only solutions to the equations describing user dynamics but also the causal relationships between the effects of specific network structures on user dynamics, we can obtain a fundamental equation that satisfies the requirements of both the solution of the wave equation and an explicit description of the causal relationships.
Interestingly, the resulting fundamental equation has a structure similar to the framework of relativistic quantum mechanics.
This paper uses this formal association with relativistic quantum mechanics to investigate the influence of a particular network structure on user dynamics based on the perturbation theory used in quantum theory.
Specifically, based on the perturbative expansion method ~\cite{aida2018generation} for analyzing the solution of the fundamental equation of an oscillation model, we consider how to evaluate the influence of the network structure on the eigenvalues by giving a perturbative expansion up to infinite order to a network model with a simple graph structure.
Numerical experiments verify the accuracy of our perturbative expansions of finite degree.

This paper is organized as follows.
In Section 2, we introduce related works and describe the position of this research.
Section 3 provides preliminary knowledge of the fundamental equation to describe the user dynamics of OSNs.
In Section 4, we apply perturbation theory to the fundamental equation of user dynamics and explain the perturbative calculation of the fundamental equation to treat the effects of some graph structures in terms of perturbation.
In Section 5, we explain our specific perturbative calculation method using a simple network model and show the results of calculating the perturbative expansion to infinite order and the method of higher-order correction.
In Section 6, we propose a method to calculate the Laplacian matrix's eigenvalues from the perturbative expansion and confirm its accuracy by numerical experiments.
In Section 7, we conclude our research. 

\section{Related Work}
In recent years, various phenomena occurring on OSNs have been modeled and assessed.
Reference \cite{pastor2001epidemic} proposed an infection model, the SIS model, on scale-free networks to investigate the spread of computer viruses.
Subsequently, the SIR model on networks was proposed~\cite{newman2002spread}.
These models were extended with the proposal of a model that deals with the diffusion of rumors~\cite{nekovee2007theory} and a model that deals with the users' adoption and abandonment of SNS ~\cite{cannarella2014epidemiological}.

Consensus problems including opinion formation of users in networks has been modeled~\cite{ren2005consensus,olfati2007consensus}.

Of significant interest, the echo chamber phenomenon, in which opinions are radicalized by repeated communication within a highly homogeneous community has recently been recognized as a problem on OSNs.
For example, reference \cite{baumann2020modeling} models the process of echo chamber generation.
Reference \cite{tornberg2018echo} clarifies the effect of the existence of echo chambers on information diffusion.

Various other network dynamics have also been modeled.
Examples include: modeling the diffusion process of innovations~\cite{iacopini2018network}, modeling the decision of different software applications~\cite{gleeson2014simple}.

Some studies have used real data to elucidate the unique characteristics of dynamics on OSNs.
These studies include information diffusion on social media~\cite{lerman2010information,bakshy2011everyone}, a mechanism of how rumors spread quickly in OSNs~\cite{doerr2012rumors}, the role of weak ties in the propagation of new information~\cite{bakshy2012role}, and the relationship between information diffusion on Twitter and the burstiness of network link generation and deletion~\cite{myers2014bursty}.
Experiments in reference \cite{centola2010spread} confirmed the characteristics of user behavior diffusion on networks with different structures.

One of the interests in such works on dynamics in networks is  clarifying the relationship between network structure and the dynamics exhibited.
For example, a study on the diffusion of user behavior~\cite{centola2010spread} revealed that a network containing a cluster structure diffuses to more people and faster than a network with a random link structure.
These models are characterized by the fact that they deal with first-order differential equations of time.

The oscillation model described in the introduction is based on the wave equation in the network, which is a second-order differential equation of time.
It assumes that the effects between users propagate over the OSN at finite speed.
In the oscillation model, the conditions under which the intensity of the user dynamics diverges can be clarified by the eigenvalues of the Laplacian matrix representing the network structure.

In this paper, based on the framework of the oscillation model, we discuss how to explicitly investigate how the user dynamics will react to changes in the network structure.

\section{Preliminary}
In this section, we outline the framework of the oscillation model for online user dynamics and the basics of the fundamental equation.

\subsection{Symmetrizable Graphs and Decomposition of Directed Graphs}
The structure of an OSN is represented by a directed graph in which users are nodes and relationships between users are directed links.
The directed graph is defined by node set $V=\{1,2,\dots,n\}$ and link set $E$; it is denoted by $\mathcal{G}(V,E)$.
Each directed link $((i \rightarrow j) \in E)$ has link weight $w_{ij}>0$.
The (weighted) adjacency matrix $\bm{\mathcal{A}} = [\mathcal{A}_{ij}]_{1\le i,j \le n}$ of directed graph $\mathcal{G}$ is defined as follows.
\begin{align}
   \mathcal{A}_{ij} := 
               \begin{cases}
                  w_{ij} \quad &( (i\rightarrow j) \in E ), \\
                  0       &( (i\rightarrow j) \notin E ).
               \end{cases}
               \label{adj}
\end{align}
Next, we define the outgoing degree $d_i$ of node $i$ as $d_{i} := \sum_{j \in \partial i} w_{ij}$ by summing the link weights $w_{ij}$ $(j\in \partial i)$ (where $\partial i$ represents the set of adjacent nodes of node $i$).
The matrix in which the outgoing degrees of each node are arranged in diagonal components is called the outgoing degree matrix and is defined as $\bm{\mathcal{D}}:= \mathrm{diag}(d_{0}, d_{1}, ... , d_{n-1})$.
The Laplacian matrix is defined as $\bm{\mathcal{L}} := \bm{\mathcal{D}}-\bm{\mathcal{A}}$ with the outgoing degree matrix $\bm{\mathcal{D}}$ and the adjacency matrix $\bm{\mathcal{A}}$.

While the Laplacian matrix of the undirected graph is a symmetric matrix, the Laplacian matrix of the directed graph is an asymmetric matrix.
Therefore, the eigenvalues of the Laplacian matrix of an undirected graph are always real and non-negative due to the nature of the Laplacian matrix.
On the other hand, the eigenvalues of the Laplacian matrix of a directed graph are not always real numbers, but the real parts of the eigenvalues are known to be non-negative~\cite{aida2020book}.

The symmetrizable graph is a directed graph in which the Laplacian matrix representing the network structure can be transformed into a symmetric matrix by similarity transformation using a diagonal matrix with positive diagonal components.
The characteristic of a symmetrizable graph is that the product of the link weights in the rightward and leftward paths of any closed path of the graph is equal.

\begin{figure}[tb].
  \centering
  \includegraphics[width=\linewidth]{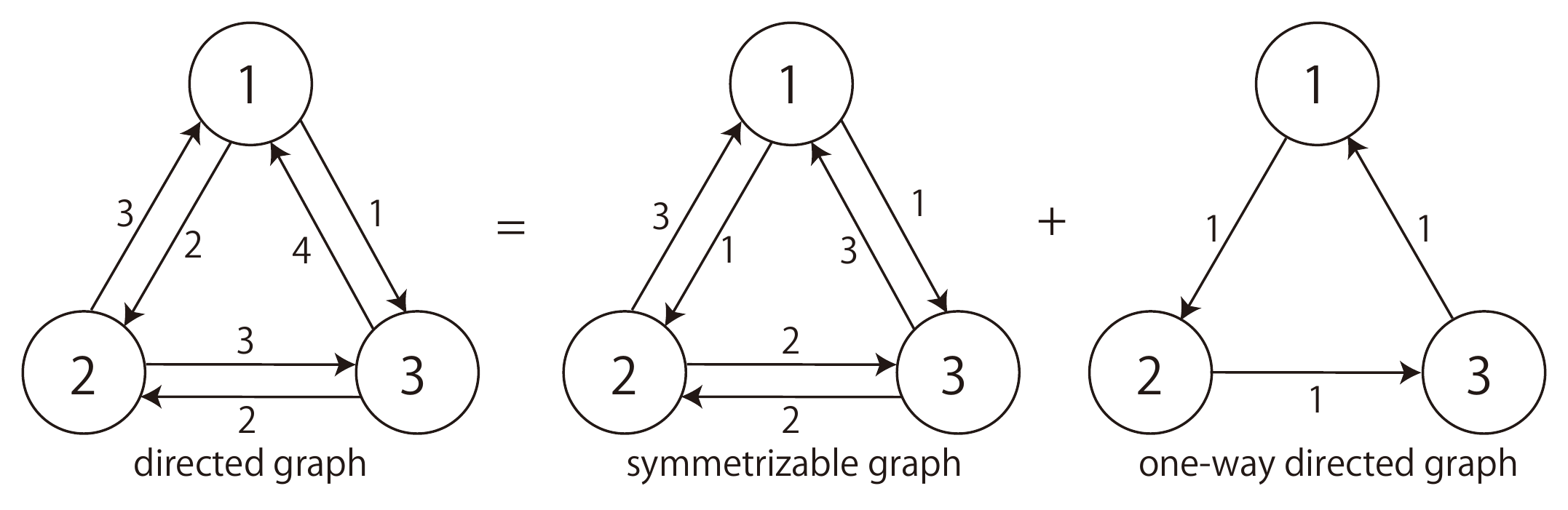}
  \caption{Decomposing a directed graph.}
  \label{sym}
\end{figure}

Any directed graph can be decomposed into a symmetrizable graph and a {\it one-way link graph} having at most only one-directional link between each node~\cite{MasakiAIDA20182017EBN0001}.
However, it is known that the decomposition is, in general, not unique.
An example of Laplacian matrix decomposition is shown in Figure~\ref{sym}.

Let the original Laplacian matrix be $\bm{\mathcal{L}}$, the symmetrizable Laplacian matrix after decomposition be $\bm{\mathcal{L}}_0$, and the Laplacian matrix corresponding to the one-way link graph be $\bm{\mathcal{L}}_\mathrm{I}$, 
\begin{align}
\bm{\mathcal{L}} &= \bm{\mathcal{L}}_0 + \bm{\mathcal{L}}_\mathrm{I}. 
\label{eq:L_decomp}
\end{align}
Then, the decomposition of Figure \ref{sym} can be written as follows:
\begin{align}
\begin{bmatrix}
3&-2&-1\\
-3&6&-3\\
-4&-2&6
\end{bmatrix}
&=
\begin{bmatrix}
2&-1&-1\\
-3&5&-2\\
-3&-2&5
\end{bmatrix}
+
\begin{bmatrix}
1&-1&0\\
0&1&-1\\
-1&0&1
\end{bmatrix}
.
\notag
\end{align} 

\subsection{Oscillation Model for Online User Dynamics}
This subsection briefly describes the oscillation model~\cite{MasakiAIDA20182017EBN0001}, a theoretical framework for modeling user dynamics in OSNs.

Consider an OSN with $n$ users, where $x_i(t)$ is the state of node (user) $i$ at time $t$.
Let the state vector of the node be $\bm{x}(t) := {}^t\! (x_1(t),\,\dots,\,x_n(t))$.
In the oscillation model in OSNs, the equation describing online user dynamics is given as follows:
\begin{align}
\frac{\dd^2}{\dd t^2} \, \bm{x}(t) = -\bm{\mathcal{L}} \, \bm{x}(t).
\label{eq:EoM}
\end{align}
This is the wave equation on the network, where $\bm{\mathcal{L}}$ is the Laplacian matrix of the directed graph representing the structure of the OSN.

From the solution of this equation, we can derive the oscillation energy; its value gives a generalized measure of node centrality.
Node centrality is the measure of the degree of importance or activity of each node in a network. Typical examples are degree centrality and betweenness centrality.
In the oscillation model, if the link weights and initial values are chosen appropriately, it is known that the oscillation energy yields degree centrality and betweenness centrality, which can provide a unifying foundation for different kinds of node centrality~\cite{TakanoIEICE.T.2018}.
Moreover, since oscillation energy represents the intensity of network activity, the phenomenon of divergence in oscillation energy under certain conditions corresponds to the phenomenon of divergence in the intensity of user activity, which can be interpreted as the occurrence of explosive user dynamics such as online flaming~\cite{aidaASONAM2017}.

The behavior of solutions to the wave equation \eqref{eq:EoM} is characterized by the eigenvalues of the Laplacian matrix, and it is known that the amplitude of the solution diverges when any of the eigenvalues are not real values.
When the amplitude of the solution diverges, the oscillation energy also diverges, and this phenomenon can be thought of as an occurrence of explosive user dynamics such as online flaming.

Thus, the conditions for the occurrence of explosive user dynamics can be explained by the eigenvalues of the Laplacian matrix representing the OSN structure.
Then, is it possible to understand the causal relationship between OSN structure and user dynamics in terms of {\it what kind of network structure causes explosive user dynamics?}
To answer this question, we divide the structure of the OSN into two parts, one is that part of the network structure whose solution properties are well known, while the other is the remainder.
In this case, it is desirable that the user dynamics of the entire OSN are given in an easily understandable form of causal relationships as expressed by the solutions to the equations obtained from the network structure with well-known properties and the solutions obtained from the remainder of the network structure.
The method is briefly summarized below.

Consider the symmetrizable Laplacian matrix $\bm{\mathcal{L}}_0$ obtained by decomposing $\bm{\mathcal{L}}$ by \eqref{eq:L_decomp}, and the Laplacian matrix of the one-way link graph $\bm{\mathcal{L}}_\mathrm{I}$.
Let $\bm{\Lambda}_0$ be the diagonalized matrix obtained from $\bm{\mathcal{L}}_0$, and let $\bm{\Lambda} = \bm{\Lambda}_0 + \bm{\Lambda}_\mathrm{I}$ be the matrix yielded by applying the same diagonalizing transformation of $\bm{\mathcal{L}}_0$ to $\bm{\mathcal{L}}$. 
Then, equation \eqref{eq:EoM} can be transformed as follows:
\begin{align}
\frac{\dd^2}{\dd t^2} \, \bm{\phi}(t) = -\bm{\Lambda} \, \bm{\phi}(t) = -(\bm{\Lambda}_0 + \bm{\Lambda}_\mathrm{I}) \, \bm{\phi}(t),
\label{eq:EoM2}
\end{align}
where $\bm{\Lambda}_\mathrm{I}$ is the matrix corresponding to the effect of the one-way link graph, and $\bm{\phi}(t)$ is the user's state vector  $\bm{x}(t)$ transformed according to the transformation of the equation.

When $\bm{\Lambda}_\mathrm{I}=\bm{\mathrm{O}}$, i.e., when $\bm{\mathcal{L}}_\mathrm{I} =\bm{\mathrm{O}}$, the transformed wave equation~\eqref{eq:EoM2} is diagonalized. 
This means we have independent $n$ wave equations. 
In this case, since all its eigenvalues of $\bm{\Lambda}$ are real (and non-negative), explosive user dynamics does not occur.
In other words, the user dynamics produced by the structure of $\bm{\mathcal{L}}_0$ corresponds to the solution of the equation whose properties are well known.
Therefore, the emergence of explosive user dynamics is driven by the one-way link graph represented by $\bm{\Lambda}_\mathrm{I}$.

Diagonalizing the wave equation for the symmetrizable directed graph $\bm{\mathcal{L}}_0$ using $\bm{\Lambda}_0$ yields
\[
\frac{\dd^2}{\dd t^2} \, \bm{\phi}_0(t) = -\bm{\Lambda}_0 \, \bm{\phi}_0(t);
\]
let $\bm{\phi}_0(t)$ be the solution to this diagonalized wave equation.
Also, let $\bm{\Phi}_0(t)$ be the $n\times n$ diagonal matrix whose diagonal components are the components of $\bm{\phi}_0(t)$.
This satisfies the same diagonalized wave equation as $\bm{\phi}_0(t)$ as follows:
\[
\frac{\dd^2}{\dd t^2} \, \bm{\Phi}_0(t) = -\bm{\Lambda}_0 \, \bm{\Phi}_0(t).
\]
Since this is $n$ independent equations, the solution is simply obtained as follows:
\[
\bm{\Phi}_0(t) := \exp\!\left(\pm\ii\sqrt{\bm{\Lambda}_0}\,t\right)  \bm{\Phi}_0(0).
\]
Next, let $\bm{\phi}_\mathrm{I}$ be the solution to the equation for the one-way link graph described below, and check whether solution $\bm{\phi}(t)$ of the transformed wave equation \eqref{eq:EoM2} can be expressed in a product-form, as an easily understood structure, as shown by 
\[
\bm{\phi}(t) = \bm{\Phi}_0(t) \, \bm{\phi}_\mathrm{I}(t). 
\]
 As an initial condition, we set $\bm{\phi}(t) = \bm{\phi}_\mathrm{I}(0)$, that is, $\bm{\Phi}_0(t) = \bm{I}$.
Considering the possibility of a product-form solution, we derive an equation with $\bm{\phi}_\mathrm{I}$ as a solution as follows: 
\[
\frac{\dd^2}{\dd t^2} \, \bm{\phi}_\mathrm{I}(t) = -\big(\bm{\Phi}_0(-t) \, \bm{\Lambda}_\mathrm{I} \, \bm{\Phi}_0(t)\Big) \bm{\phi}_\mathrm{I}(t). 
\]
 
Using the fact that $\bm{\Phi}_0(-t) = (\bm{\Phi}_0(t))^{-1}$, we obtain
\begin{align}
\frac{\dd^2}{\dd t^2} \, \bm{\phi}(t) &= \frac{\dd^2}{\dd t^2} \, (\bm{\Phi}_0(t) \, \bm{\phi}_\mathrm{I}(t))
\notag\\
&= \frac{\dd}{\dd t} \left(\frac{\dd \bm{\Phi}_0(t)}{\dd t} \, \bm{\phi}_\mathrm{I}(t) + \bm{\Phi}_0(t) \, \frac{\dd \bm{\phi}_\mathrm{I}(t)}{\dd t}\right)
\notag\\
&= \Big(\frac{\dd^2 \bm{\Phi}_0(t)}{\dd t^2} \, \bm{\phi}_\mathrm{I}(t) + \bm{\Phi}_0(t) \, \frac{\dd^2 \bm{\phi}_\mathrm{I}(t)}{\dd t^2} 
\notag\\
& \qquad\qquad\qquad {}+ 2\,\frac{\dd \bm{\Phi}_0(t)}{\dd t} \,\frac{\dd \bm{\phi}_\mathrm{I}(t)}{\dd t}\Big)
\notag\\
&= -(\bm{\Lambda}_0 + \bm{\Lambda}_\mathrm{I})\,\bm{\phi}(t) + + 2\,\frac{\dd \bm{\Phi}_0(t)}{\dd t} \,\frac{\dd \bm{\phi}_\mathrm{I}(t)}{\dd t}
\notag\\
&\not= -\bm{\Lambda}\,\bm{\phi}(t). 
\end{align}
Since the last equality does not hold, the attempt to obtain a product-form solution is not successful.
This is because the wave equation is a second-order differential equation of time, which results in an extra cross-term.
To solve this problem, we need to rewrite the wave equation as a first-order differential equation of time.
At the same time, the solution to the rewritten first-order differential equation must also be a solution to the original second-order differential wave equation.

We solve this problem by introducing the following fundamental equation of the oscillation model:
\begin{align}
  \pm \ii\,\frac{\dd}{\dd t}\,\bm{\psi}(t)
  = \bm{\Omega} \, \bm{\psi}(t) = \big(\bm{\Omega}_0+\bm{\Omega}_\mathrm{I}\big) \, \bm{\psi}(t),
  \label{fundamental_eq}
\end{align}
where $\bm{\Omega}_0$ and $\bm{\Omega}_\mathrm{I}$ are $n\times n$ matrixes determined from $\bm{\Omega}_0^2=\bm{\Lambda}_0$,  $(\bm{\Omega}_0+\bm{\Omega}_\mathrm{I})^2=\bm{\Lambda}$.
$\bm{\Omega}_\mathrm{I}$ is the matrix that represents the effect of the one-way link graph and is the matrix that can be the cause of the explosive user dynamics.
The solution can then be given as a product-form solution as follows:
\begin{align}
  \bm{\psi}(t) = \bm{\Psi}_0(t) \, \bm{\psi}_\mathrm{I}(t).
\label{product_form}
\end{align}
Note that $\bm{\Psi}_0(t)$ and $\bm{\psi}_I(t)$ are solutions to the following equation:
\begin{align}
  \pm\ii \, \frac{\dd}{\dd t} \, \bm{\Psi}_0(t) &=
  \bm{\Omega}_0\,\bm{\Psi}_0(t)
  \label{Phi_0}\\
  \pm\ii \, \frac{\dd}{\dd t} \, \bm{\psi}_\mathrm{I}(t) &=
  \Big(\bm{\Psi}_0(-t) \, \bm{\Omega}_\mathrm{I} \, \bm{\Psi}_0(t)\Big) \, \bm{\psi}_\mathrm{I}(t), 
  \label{Phi_I}
\end{align}
where $\bm{\Psi}_0(t)$ is an $n\times n$ matrix with $\bm{\Psi}_0(0) = \bm{I}$.
This is checked below.
Substituting the product-form solution (\ref{product_form}) into the fundamental equation (\ref{fundamental_eq}), we get 
\begin{align}
  \pm \ii\,\frac{\dd}{\dd t}\,\bm{\psi}(t)
  &= \pm \ii\,\frac{\dd}{\dd t}\,\left(\bm{\Psi}_0(t) \, \bm{\psi}_\mathrm{I}(t)\right)
  \notag\\
  &= \pm \ii\,\frac{\dd \bm{\Psi}_0(t)}{\dd t}\,\bm{\psi}_\mathrm{I}(t)
  \pm \ii\,\bm{\Psi}_0(t) \,\frac{\dd \bm{\psi}_\mathrm{I}(t)}{\dd t}
  \notag\\
  &= \bm{\Omega}_0\,\bm{\Psi}_0(t)\,\bm{\psi}_\mathrm{I}(t)
  \notag\\
  &\qquad +\bm{\Psi}_0(t) \,  \Big(\bm{\Psi}_0(-t) \, \bm{\Omega}_\mathrm{I} \, \bm{\Psi}_0(t)\Big) \, \bm{\psi}_\mathrm{I}(t)
  \notag\\
  &=\big(\bm{\Omega}_0+\bm{\Omega}_\mathrm{I}\big) \, (\bm{\Psi}_0(t)\,\bm{\psi}(t))
  \notag\\
  &= \bm{\Omega} \,\bm{\psi}(t), 
\end{align}
and the product-form solution is the solution to the fundamental equation.
Note that we used the fact that $\bm{\Psi}_0(-t)=(\bm{\Psi}_0(t))^{-1}$. 

Solution $\bm{\psi}(t)$ of fundamental equation~(\ref{fundamental_eq}) also satisfies equation~(\ref{eq:EoM2}).
This can be checked using (\ref{fundamental_eq}) as follows: 
\begin{align}
\frac{\dd^2}{\dd t^2} \, \bm{\psi}(t) &= \mp\ii\frac{\dd}{\dd t}\,\bm{\Omega}\, \bm{\psi}(t) = -\bm{\Omega}^2 \, \bm{\psi}(t) = -\bm{\mathcal{L}} \, \bm{\psi}(t) .
\end{align}

\section{Description of Causal Relationship Between OSN Structure and User Dynamics Using Perturbation Theory}
\subsection{Perturbative Expansion of Fundamental Equation}
\label{expansion_section}
Since the property of the fundamental equation (\ref{fundamental_eq}) is that the influence of the network structure on the solution can be described by the product-form solution, we use perturbation theory to explicitly describe the influence of the one-way link graph when the Laplacian matrix of OSN is decomposed, see (\ref{eq:L_decomp}).
Perturbation theory is an approach that finds approximate solutions by slightly changing a simple equation whose solution has well-known properties.
As the equation and its solution with well-known properties are (\ref{Phi_0}) and $\bm{\Phi}_0(t)$, respectively,  the parameter that characterizes the strength of the influence of the one-way link graph is $\epsilon$. It is defined as 
\[
\bm{\Omega}(\epsilon) := \bm{\Omega}_0+\epsilon\,\bm{\Omega}_\mathrm{I}. 
\]
Depending on the value of $\epsilon$, $\bm{\Omega}(0) = \bm{\Omega}_0$, $\bm{\Omega}(1) = \bm{\Omega}$.
Next, consider the fundamental equation with $\bm{\Omega}(\epsilon)$ as follows: 
\begin{align}
  \pm \ii \, \frac{\dd}{\dd t} \, \bm{\psi}(\epsilon;t)
  = \bm{\Omega}(\epsilon) \, \bm{\psi}(\epsilon;t) = (\bm{\Omega}_0+\epsilon\,\bm{\Omega}_\mathrm{I}) \, \bm{\psi}(\epsilon;t), 
  \label{fundamental_eq_perturbation}
\end{align}
where the solution to this equation, $\bm{\psi}(\epsilon;t)$, is a vector that depends on parameter $\epsilon$, where $\bm{\psi}(0;t) = \bm{\psi}_0(t)$ and $\bm{\psi}(1;t) = \bm{\psi}(t)$.
In perturbation theory, we investigate how $\bm{\psi}(\epsilon;t)$ for $\epsilon \not = 0$ changes with respect to $\bm{\psi}_0(t)$ as a power series of $\epsilon$.
The method is specified as follows.

The formal solution to (\ref{fundamental_eq_perturbation}) is $\bm{\psi}(t) = \exp(\mp\bm{\Omega}(\epsilon))\, \bm{\psi}(0)$, but $\bm{\Omega}(\epsilon) $ is not a diagonal matrix for $\epsilon>0$, so the solution cannot be expressed simply.
For this reason, we introduce a procedure that offers expansion as a power series of $\epsilon$ as follows.
First, for $\Delta t \ll 1$, if $t = N\,\Delta t$, we get 
\begin{align}
  \bm{\psi(t)} &= (1\mp \ii\,\bm{\Omega}(\epsilon)\,\Delta t)^N \, \bm{\psi}(0)\notag\\
  &=\left((1\mp \ii\,\bm{\Omega}_0\,\Delta t)+(\mp \ii\,\epsilon\,\bm{\Omega}_\mathrm{I}\,\Delta t)\right)^N \, \bm{\psi}(0). 
  \label{sol_discrete}
\end{align}
Next, the perturbative expansion of solution $\bm{\psi}(t)$ with $\epsilon$ is as follows: 
\begin{align}
  \bm{\psi}(t)=\bm{\psi}^{(0)}(t)+\epsilon\,\bm{\psi}^{(1)}(t)+\epsilon^2\,\bm{\psi}^{(2)}(t)+\cdots. 
  \label{expansion}
\end{align}
Consider a solution of \eqref{expansion} for each power order of $\epsilon$.
Since term $\epsilon^0$ is the case where $\bm{\Omega}_\mathrm{I}$ is not affected until time $t$, it corresponds to the term without $\epsilon$ in the binomial expansion of (\ref{sol_discrete}), so
\begin{align*}
  \bm{\psi}^{(0)}(t) = (1\mp\ii\, \bm{\Omega}_0\,\Delta t)^N\, \bm{\psi}(0), 
\end{align*}
and when $\Delta t\rightarrow 0$, we get 
\begin{align*}
  \bm{\psi}^{(0)}(t) = \exp(\mp \ii \, \bm{\Omega}_0 \, t) \, \bm{\psi}(0).
\end{align*}
The term of $\epsilon^1$, when there is an effect of $\bm{\Omega}_\mathrm{I}$ at time $t_1 = k_1\,\Delta t$, is written as
\begin{align*}
  &(1\mp \ii\,\bm{\Omega}_0\,\Delta t)^{N-k_2} \,
  (\mp \ii \, \bm{\Omega}_\mathrm{I} \, \Delta t) \,
  (1\mp \ii \, \bm{\Omega}_0 \, \Delta t)^{k_2-k_1-1}
\notag
\\
&\times(\mp \ii \, \bm{\Omega}_\mathrm{I}\Delta t) \,
  (1\mp \ii \, \bm{\Omega}_0 \, \Delta t)^{k_1-1} \,
  \bm{\psi}(0); 
\end{align*}
adding them up gives the difference at time $t_1$. 
When $\Delta t\rightarrow 0$, we get 
\begin{align*}
  &\bm{\psi}^{(2)}(t)=
  \Bigg(\int_{0}^{t} \!\!\! \int_{t_1}^{t}
  \e^{\mp \ii \, \bm{\Omega}_0 \, (t-t_2)} \,
  (\mp\ii \, \bm{\Omega}_\mathrm{I}) \,
  \e^{\mp\ii \, \bm{\Omega}_0 \, (t_2-t_1)}
\notag
\\
&\qquad\qquad\qquad\qquad
    \times(\mp\ii \, \bm{\Omega}_\mathrm{I}) \,
  \e^{\mp\ii \, \bm{\Omega}_0 \, t_1} \,
  \dd t_2 \, \dd t_1\Bigg) \, 
  \bm{\psi}(0). 
\end{align*}
The term of $\epsilon^2$, when affected by $\bm{\Omega}_\mathrm{I}$ at times $t_1=k_1\,\Delta t$ and $t_2=k_2\,\Delta t$ $(t_1 < t_2)$, is 
\begin{align*}
  &(1\mp \ii\,\bm{\Omega}_0\,\Delta t)^{N-k_2} \,
  (\mp \ii \, \bm{\Omega}_\mathrm{I} \, \Delta t) \,
  (1\mp \ii \, \bm{\Omega}_0 \, \Delta t)^{k_2-k_1-1}
\notag
\\
&\times(\mp \ii \, \bm{\Omega}_\mathrm{I}\Delta t) \,
  (1\mp \ii \, \bm{\Omega}_0 \, \Delta t)^{k_1-1} \,
  \bm{\psi}(0);
\end{align*}
adding them up gives the difference between time $t_1$ and $t_2$.
When $\Delta t\rightarrow 0$, we get 
\begin{align*}
  &\bm{\psi}^{(2)}(t)=
  \Bigg(\int_{0}^{t} \!\!\! \int_{t_1}^{t}
  \e^{\mp \ii \, \bm{\Omega}_0 \, (t-t_2)} \,
  (\mp\ii \, \bm{\Omega}_\mathrm{I}) \,
  \e^{\mp\ii \, \bm{\Omega}_0 \, (t_2-t_1)}
\notag
\\
&\qquad\qquad\qquad\qquad
    \times(\mp\ii \, \bm{\Omega}_\mathrm{I}) \,
  \e^{\mp\ii \, \bm{\Omega}_0 \, t_1} \,
  \dd t_2 \, \dd t_1\Bigg) \, 
  \bm{\psi}(0). 
\end{align*}
For higher-order terms, we can generate the integrand in the same way and calculate $k$ multiple integrals to obtain the solution for $\epsilon^k$ order.

\subsection{Model of Coupled Oscillation Modes and Example of Perturbative Expansion}
\label{formulation_section}
Hereafter, since both equations in \eqref{fundamental_eq} have the same structure, we examine the fundamental equation of 
\begin{align*}
  + \ii\,\frac{\dd}{\dd t}\,\bm{\psi}(t)
  = \bm{\Omega} \, \bm{\psi}(t) = \big(\bm{\Omega}_0+\bm{\Omega}_\mathrm{I}\big) \, \bm{\psi}(t),
\end{align*}
from the two equations in \eqref{fundamental_eq}.
 
\begin{figure}[tb]
  \centering
  \includegraphics[width=0.5\linewidth]{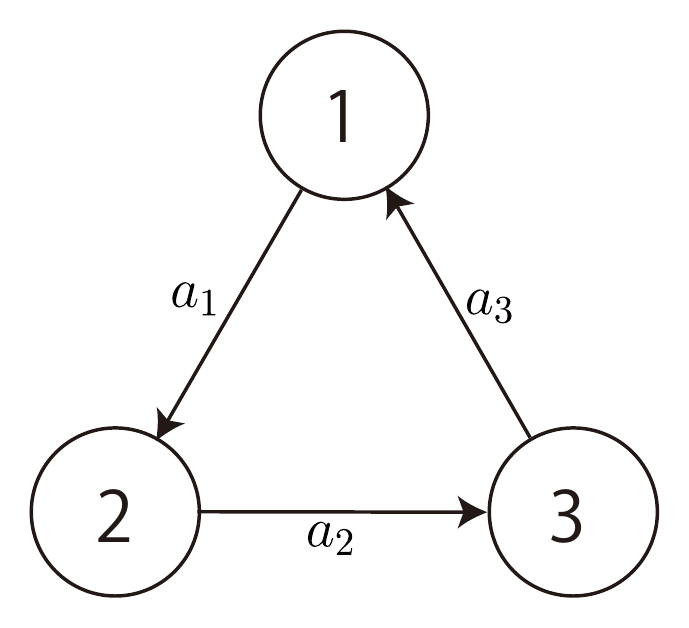}
  \caption{Cyclic coupling relation of three oscillation modes.}
  \label{mode}
\end{figure}

There are $n$ oscillation modes that emerge from the wave equation representing the user dynamics for a network with $n$ users. To understand the mechanism of perturbative expansion, we consider a model in which only three of the $n$ oscillation modes affect each other cyclically, as shown in Figure~\ref{mode}.
Omitting the independent oscillation modes, we focus only on the $3$ oscillation modes that affect each other as represented by the $3\times 3$ matrix, $\bm{\Omega}(\epsilon)$, as follows: 
\begin{align}
  \bm{\Omega}(\epsilon)&=\bm{\Omega}_0+\epsilon\,\bm{\Omega}_\mathrm{I}
  \notag\\
  &=
  \begin{bmatrix}
    \omega_1&0&0\\
    0&\omega_2&0\\
    0&0&\omega_3
  \end{bmatrix}
  +\epsilon
  \begin{bmatrix}
    d_1&-a_1&0\\
    0&d_2&-a_2\\
    -a_3&0&d_3
  \end{bmatrix}
  .
\end{align}
Now, we decompose $\bm{\Omega}(\epsilon)$ into a diagonal matrix and the remainder as follows: \begin{align}
  \bm{\Omega}(\epsilon) &=
  \begin{bmatrix}
    \omega_1+\epsilon d_1&0&0\\
    0&\omega_2+\epsilon d_2&0\\
    0&0&\omega_3+\epsilon d_3
  \end{bmatrix}
  \notag
  \\
  &\quad
  +\epsilon
  \begin{bmatrix}
    0&-a_1&0\\
    0&0&-a_2\\
    -a_3&0&0
  \end{bmatrix}
  \notag\\
  &=\begin{bmatrix}
    \omega_1^{\prime}&0&0\\
    0&\omega_2^{\prime}&0\\
    0&0&\omega_3^{\prime}
  \end{bmatrix}
  +\epsilon \,
  \begin{bmatrix}
    0&-a_1&0\\
    0&0&-a_2\\
    -a_3&0&0
  \end{bmatrix}
  . 
  \label{Omega}
\end{align}

\begin{figure}[tb]
  \centering
  \includegraphics[width=\linewidth]{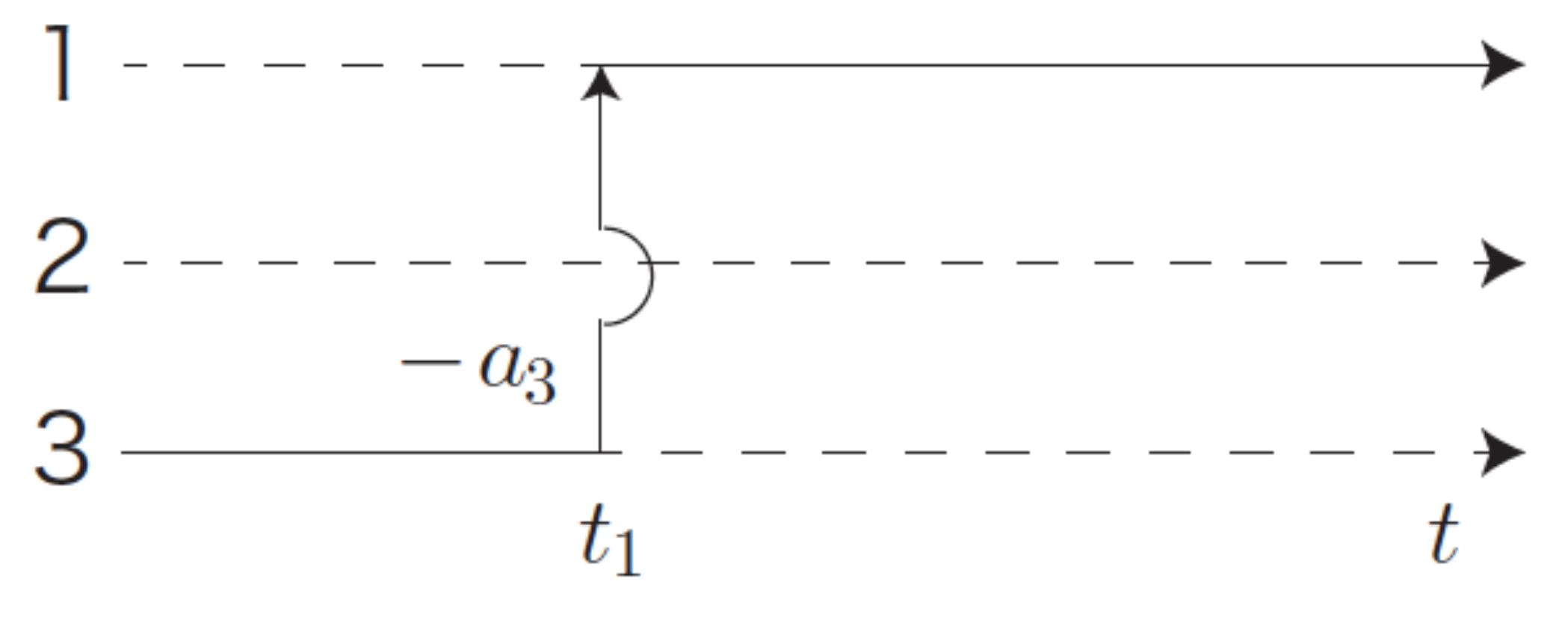}
  \caption{State transition diagram for $\epsilon^1$ order.}
  \label{first_order}
\end{figure}

We set the initial state as follows: 
\begin{align*}
  \bm{\psi}(0) = {}^t\!(\psi_1(0), \, \psi_2(0), \, \psi_3(0)), 
\end{align*}
and consider $\psi_1(t)$ of node 1 on behalf of the state of nodes.
The perturbative expansion by the effect of the non-diagonal component of $\epsilon\,\bm{\Omega}_\mathrm{I}$ is as follows: 
\begin{align}
\psi_1(t)=\psi_1^{(0)}(t)+\epsilon\,\psi_1^{(1)}(t)+\epsilon^2\,\psi_1^{(2)}(t)+\cdots. 
\label{psi_1}
\end{align}
Let us consider the contribution of each term for each order of $\epsilon$.
Since there is no effect of $\bm{\Omega}_\mathrm{I}$, the contribution of the $\epsilon^0$ order is 
\begin{align}
  \psi_1^{(0)}(t)=\exp(-\ii\,\omega_1^{\prime} t) \, \psi_1(0).
\end{align}
Next, we consider the contribution of the $\epsilon^1$ order. 
Figure~\ref{first_order} shows the state transition corresponding to the contribution of the $\epsilon^1$ order; this type of diagram is called the Feynman diagram. 

The contribution of the $\epsilon^1$ order is calculated as follows: 
\begin{align}
   \psi_1^{(1)}(t) &= \left(\int_0^t
  \e^{-\ii\,\omega_1^{\prime} (t-t_1)}\,
  (-\ii \,\epsilon\,a_3)\,
  \e^{-\ii\,\omega_3^{\prime} t_1}\,
  \dd t_1\right) \psi_3(0)
  \notag\\
  &= \frac{\epsilon\,a_3\, ( \e^{-\ii \omega _1^{\prime} t} - \e^{-\ii \omega _3^{\prime} t} ) }{\omega _3^{\prime}-\omega _1^{\prime}}\,\psi_3(0). 
  \label{eq:first_order}
\end{align}

\begin{figure}[tb]
  \centering
  \includegraphics[width=\linewidth]{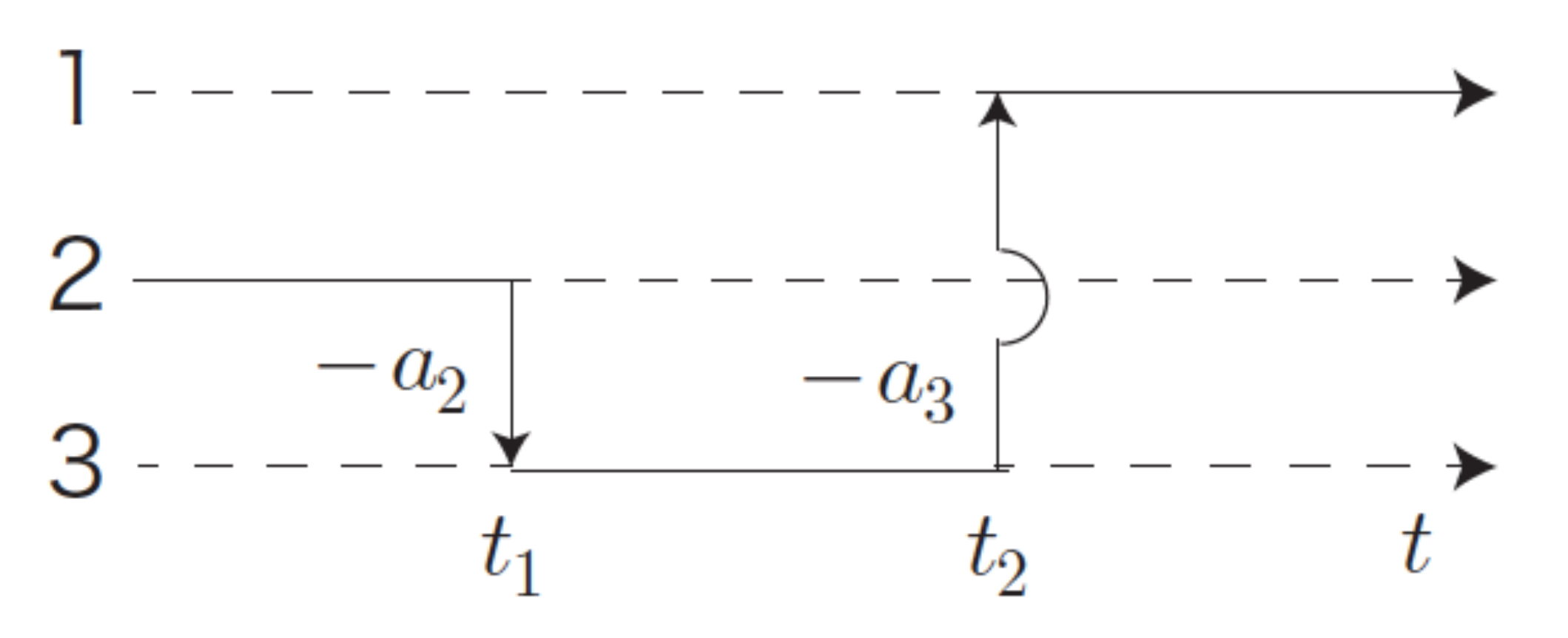}
  \caption{State transition diagram for $\epsilon^2$ order.}
  \label{second_order}
\end{figure}

Figure~\ref{second_order} shows the Feynman diagram corresponding to the contribution of the $\epsilon^2$ order. 

The contribution of the $\epsilon^2$ order is calculated as follows: 
\begin{align}
   &\psi_1^{(2)}(t) 
   \notag\\
   &= \Bigg(\int_0^t \!\!\!
  \int_{t_1}^t
  \e^{-\ii\,\omega_1^{\prime} (t-t_2)}\,
  (-\ii \, \epsilon\,a_3)\,
  \e^{-\ii\,\omega_3^{\prime} (t_2-t_1)}\notag\\
  &\qquad \times(-\ii \, \epsilon\,a_2)\,
  \e^{-\ii\,\omega_2^{\prime}\,t_1}\,
  \dd t_2 \,
  \dd t_1\Bigg) \, \psi_2(0)
  \notag\\
  &= \frac{\epsilon^2\,a_2\,a_3}{\omega^{\prime}_3-\omega^{\prime}_1}  \left(\frac{\e^{-\ii {\omega^{\prime}_2}\,t} - \e^{-\ii
   \omega^{\prime}_3\,t}}{\omega^{\prime}_2-\omega^{\prime}_3}-\frac{\e^{-\ii 
   \omega^{\prime}_1\,t} - \e^{-\ii  \omega^{\prime}_2\,t}}{\omega^{\prime}_1-\omega^{\prime}_2}\right)\,\psi_2(0). 
  \label{eq:second_order}
\end{align}
As shown above, we can consider the contribution of each order of $\epsilon$, and by calculating the contribution to higher orders and finding regularities, we can evaluate the perturbative expansion with regard to infinite orders.

\section{Perturbative Expansion Up to Infinite Order}
\label{infinite}
We compute higher-order perturbations on the model of cyclic coupling of the $3$ oscillation modes shown in Sect.~\ref{formulation_section} and infer the perturbative expansion to infinite order from the regularities appearing in the perturbative expansion results.

\subsection{Inference of Regularities of Perturbative Expansion}
As shown in \eqref{eq:first_order} and \eqref{eq:second_order}, the results of the perturbative expansion become more complex as the order increases.
However, by applying an appropriate representation, the perturbative calculation of $\epsilon^n$ order for $n$ can be summarized as follows: 
\begin{align}
  &\psi_1^{(n)}(t) 
  \notag\\
  &=\left( \alpha_n\, \e^{-\ii \, \omega_1^{\prime} t}
  +\beta_n\, \e^{-\ii \, \omega_3^{\prime} t}
  +\gamma_n\, \e^{-\ii \, \omega_2^{\prime} t}
  \right)\notag\\
  &\quad\times 
  \epsilon^n\,a_1{}^{[n/3]}\,a_2{}^{[(n+1)/3]}\,a_3{}^{[(n+2)/3]}
  \,\psi_\mu(0), 
  \label{deformation_step0}
\end{align}
where $\alpha_n$, $\beta_n$ and $\gamma_n$ are coefficients, and $[x]$ is the integer part of $x$. 
The index $\mu$ of $\psi_\mu(0)$ is $\mu=1$ for $n=3k$, $\mu=3$ for $n=3k+1$, and $\mu=2$ for $n=3k+2$. 
This form makes it easy to compare the results of perturbative calculations for different orders, and the regularity found through this form allows us to derive a perturbative expansion that takes account of up to infinite orders.

First, we describe the details of the transformation to the form of \eqref{deformation_step0} using the result of the $\epsilon^3$ order perturbative calculation as an example.
The contribution of the $\epsilon^3$ order is calculated according to the method in Sect.~\ref{expansion_section} as follows: \begin{align}
  \psi_1^{(3)}(t) &=
  \Bigg(\int_0^t \!\! \int_{t_1}^{t} \!\! \int_{t_2}^{t}
  \e^{-\ii\,\omega_1^{\prime} (t-t_3)} \, 
  (-\ii\,\epsilon\,a_3)\,
  \notag
  \\
  &
  \qquad\times \e^{-\ii \, \omega_3^{\prime} (t_3-t_2)}
  (-\ii \, \epsilon\,a_2) \, 
  \e^{-\ii \, \omega_2^{\prime} (t_2-t_1)}
  \notag
  \\
  &
  \qquad\times
  (-\ii \, \epsilon\,a_1)
  \e^{-\ii \, \omega_1^{\prime} t_1}
  \dd t_3 \, \dd t_2 \, \dd t_1 \Bigg) \, \psi_1(0)
  \notag\\
  &=
  \frac{\epsilon^3\,a_1 \, a_2 \, a_3}{(\omega _1^{\prime }-\omega_3^{\prime }){}^2} 
  \notag\\
  &\quad\times\Biggl(\frac{\e^{-\ii \omega _1^{\prime }\,t}-\e^{-\ii \omega _3^{\prime
   }\,t}}{\omega _2^{\prime}-\omega _3^{\prime}}
   -\frac{\ii t \left(\omega _1^{\prime}-\omega_3^{\prime }\right) \, \e^{-\ii \omega _1^{\prime}\,t}}{\omega _1^{\prime}-\omega _2^{\prime }}
   \notag\\
   &\quad\quad
   -\frac{\left(\omega _1^{\prime }-\omega _3^{\prime }\right)
   \left(\e^{-\ii \omega _1^{\prime }\,t}-\e^{-\ii \omega _2^{\prime
   }\,t}\right)}{(\omega _1^{\prime }-\omega _2^{\prime})^2}
   \notag\\
   &\quad\quad
   -\frac{(\omega _1^{\prime}-\omega_3^{\prime })\left(\e^{-\ii \omega_1^{\prime}\,t}-\e^{-\ii \omega _2^{\prime}\,t}\right)}{(\omega_1^{\prime }-\omega _2^{\prime }) \, (\omega_2^{\prime }-\omega _3^{\prime })}\Biggr) \, \psi_1(0). 
  \label{3-order_perturbation}
\end{align}
By organizing \eqref{3-order_perturbation} in the form of \eqref{deformation_step0}, the parts other than $\psi_1(0)$ are as follows: 
\begin{align}
&\!\!\!\!\!\!\!
\Biggl(
  \alpha_0\, \e^{-\ii \omega_1^{\prime} t}
  +
  \alpha_1\, \e^{-\ii \omega_1^{\prime} t}
  +
  \beta_0\, \e^{-\ii \omega_3^{\prime} t}
  +
  \gamma_0\, \e^{-\ii \omega_2^{\prime} t}
  \Biggr)\, a_1\,a_2\,a_3\,\epsilon^3
  \notag\\
  &=
  \frac{\left(2 \omega_1^{\prime}-\omega_2^{\prime}-\omega_3^{\prime}\right) a_1\,a_2\,a_3\,\epsilon^3 \,\e^{-\ii \omega_1^{\prime} t}}{\left(\omega_1^{\prime}-\omega_2^{\prime}\right){}^2 \, (\omega_3^{\prime}-\omega_1^{\prime})^2}
  \notag\\
  &\quad{}-\frac{\ii\,t\,a_1\,a_2\,a_3\,\epsilon^3 \,\e^{-\ii \omega_1^{\prime} t}}{(\omega_1^{\prime}-\omega_2^{\prime})\, (\omega_3^{\prime}-\omega_1^{\prime})}
  -\frac{a_1\,a_2\,a_3\,\epsilon^3 \,\e^{-\ii \omega_2^{\prime} t}}{(\omega_1^{\prime}-\omega_2^{\prime})^2 \,  (\omega_2^{\prime}-\omega_3^{\prime})}
  \notag\\
  &\quad{}
  +\frac{a_1\,a_2\,a_3\,\epsilon^3 \,\e^{-\ii \omega_3^{\prime} t}}{(\omega_3^{\prime}-\omega_1^{\prime})^2 \,(\omega_2^{\prime}-\omega_3^{\prime})}. 
  \label{deformation_step1}
\end{align}
Next, we transform the numerator so that it does not contain $\omega_1'$, $\omega_2'$, and $\omega_3'$ (except for the exponent of the exponential function). 
Focusing on the first term of expression~\eqref{deformation_step1}, the numerator $\left(2\omega_1^{\prime}-\omega_2^{\prime}-\omega_3^{\prime}\right)$ can be transformed as follows: 
\[
(\omega_1^{\prime}-\omega_2^{\prime})-(\omega_3^{\prime}-\omega_1^{\prime}). 
\]
These terms are canceled by the denominator and are as follows: 
\begin{align}
  \frac{a_1\,a_2\,a_3\,\epsilon^3 \e^{-\ii \, \omega_1^{\prime}\,t}}{(\omega_1^{\prime}-\omega_2^{\prime}) \, (\omega_3^{\prime}-\omega_1^{\prime})^2}
  -\frac{a_1\,a_2\,a_3\,\epsilon^3 \e^{-\ii \, \omega_1^{\prime}\,t}}{(\omega_1^{\prime}-\omega_2^{\prime})^2 \,  (\omega_3^{\prime}-\omega_1^{\prime})}. 
  \label{deformation_step2}
\end{align}
This shows that the result of the third-order perturbation can be expressed in the form of (\ref{deformation_step0}).
By transforming the numerator so as not to include $\omega_1'$, $\omega_2'$, and $\omega_3'$, it becomes easier to compare the results of different orders.

So far, we have used third-order perturbations as an example. 
For higher-order perturbation contributions, higher-order $\omega_1^{\prime}$, $\omega_2^{\prime}$, and $\omega_3^{\prime}$ appear in the numerator. 
However, it is possible to remove $\omega_1'$, $\omega_2'$, and $\omega_3'$ from the numerator by the same operation. 
To check this, we start by considering how $\alpha_m$ can be written for arbitrary $m$.
Classifying the order of perturbation $n$ into the $3$ moduli, we consider the case where $n = 3\,k$ (where $k$ denotes the number of state transitions). 
The cases where $n=3k+1$ and $n=3k+2$ are discussed later.  
Let the denominator of $\alpha_n$ be $(\omega_1^{\prime}-\omega_2^{\prime})^{n-k} \, (\omega_3^{\prime}-\omega_1^{\prime})^{n-k}$. 
Organizing the numerator by terms proportional to $(\ii\,t)^m$ $(m=0,\,1,\,\dots,\,k)$, we obtain \begin{align}
  \alpha_n &= \frac{\displaystyle \sum_{m=0}^{k}(\ii \,t)^{m}\sum_{l=0}^{k-m} r^{(\alpha)}_{lk} \, (\omega_1^{\prime}-\omega_2^{\prime})^{k-l}\,(\omega_3^{\prime}-\omega_1^{\prime})^{l+m}}{(\omega_1^{\prime}-\omega_2^{\prime})^{n-k}\,(\omega_3^{\prime}-\omega_1^{\prime})^{n-k}}, 
  \label{deformation_step3}
\end{align}
where $r^{(\alpha)}_{lk}$ is a constant determined for each combination of $l$ and $k$. 
Similarly, 
\begin{align}
  \beta_n &= \frac{\displaystyle \sum_{m=0}^{k}(\ii \,t)^{m}\sum_{l=0}^{k-m-1} r^{(\beta)}_{lk} \, (\omega_2^{\prime}-\omega_3^{\prime})^{k-l}\,(\omega_3^{\prime}-\omega_1^{\prime})^{l+m}}{(\omega_2^{\prime}-\omega_3^{\prime})^{n-k}\,(\omega_3^{\prime}-\omega_1^{\prime})^{n-k}}
\label{deformation_step3-2}\\
  \gamma_n &= \frac{\displaystyle \sum_{m=0}^{k}(\ii \,t)^{m}\sum_{l=0}^{k-m-1} r^{(\gamma)}_{lk} \, (\omega_2^{\prime}-\omega_3^{\prime})^{k-l}\,(\omega_1^{\prime}-\omega_2^{\prime})^{l+m}}{(\omega_2^{\prime}-\omega_3^{\prime})^{n-k}\,(\omega_1^{\prime}-\omega_2^{\prime})^{n-k}}, 
\label{deformation_step3-3}
\end{align}
where $r^{(\beta)}_{lk}$ and $r^{(\gamma)}_{lk}$ are constants determined for each $l$ and $k$, respectively, and the empty sum is defined as $0$.

In \eqref{deformation_step3}--\eqref{deformation_step3-3}, each term in the numerator can be reduced. 
Only the constants $r^{(\alpha)}_{lk}$, $r^{(\beta)}_{lk}$, or $r^{(\gamma)}_{lk}$ and $(\ii t)^m$ remain in the numerator of each fraction after the reduction. 
Since expression \eqref{deformation_step2} is in reduced form and corresponds to the part $k=1$, $m=0$ of expression \eqref{deformation_step3}, it can be written as two terms, $l=0$ and $l=k-m=1$. 
After transforming the numerator into a constant, each term is uniquely determined from the number of cycles of the state transition $k$, $l$ in expression \eqref{deformation_step3}, and the power $m$ of $\ii t$. 
Accordingly, in the following, each term of $\alpha_n$ in expression \eqref{deformation_step0}, after transforming it so as not to include $\omega_1'$, $\omega_2'$, and $\omega_3'$ in the numerator, is denoted as $T^{(n)}_{\alpha}(k,l,m)$. 
That is, it is defined as follows: 
\begin{align}
  &T^{(n)}_{\alpha}(k,l,m) 
  \notag\\
  &:= \frac{(\ii \,t)^{m}\,r^{(\alpha)}_{lk} \, (\omega_1^{\prime}-\omega_2^{\prime})^{k-l}\,(\omega_3^{\prime}-\omega_1^{\prime})^{l+m}}{(\omega_1^{\prime}-\omega_2^{\prime})^{n-k}\,(\omega_3^{\prime}-\omega_1^{\prime})^{n-k}}
  \notag\\
  &\quad{}\times \epsilon^n\,a_1{}^{[n/3]}\,a_2{}^{[(n+1)/3]}\,a_3{}^{[(n+2)/3]}\,\e^{-\ii\,\omega_1'\,t}. 
\end{align}
We denote $\beta_n$ and $\gamma_n$ as $T^{(n)}_{\beta}(k,l,m)$ and $T^{(n)}_{\gamma}(k,l,m)$, respectively.
Using these terms, \eqref{deformation_step0} can be written as follows: 
\begin{align}
  &\psi_1^{(n)}(t)
  \notag\\
  &=\sum_{m=0}^{k}\sum_{l=0}^{k-m} T^{(n)}_{\alpha}(k,l,m)\, \psi_\mu(0)
  \notag\\
  &\quad{}+\sum_{m=0}^{k}\sum_{l=0}^{k-m-1} \left(T^{(n)}_{\beta}(k,l,m)+T^{(n)}_{\gamma}(k,l,m)\right)\, \psi_\mu(0). 
\end{align}

Next, we explain regularities found in the results obtained by the transformation so that their numerators do not include $\omega_1'$, $\omega_2'$, or $\omega_3'$, and generalize to an infinite order perturbative expansion. 

In the following, we show expressions (\ref{deformation_step0}) for the third-, sixth-, and ninth-order perturbations and give details of regularities that can be found there.
First, for the third-order perturbation, to simplify the expression, the quantity divided by $\psi_1(0)$ is shown as follows: 
\begin{align}
&\frac{\psi_1^{(3)}(t)}{\psi_1(0)}
  \notag\\
  &= T_{\alpha}(1,0,0) + T_{\alpha}(1,1,0) + T_{\alpha}(1,0,1) + T_{\beta}(1,0,0)\notag\\
    &{}+ T_{\gamma}(1,0,0)\notag\\
    &=
    \frac{a_1\,a_2\,a_3\,\epsilon^3 \,\e^{-\ii \, \omega_1^{\prime}\,t}}
    {(\omega_1^{\prime}-\omega_2^{\prime})\, (\omega_3^{\prime}-\omega_1^{\prime})^2}
    -\frac{a_1\,a_2\,a_3\,\epsilon^3 \,\e^{-\ii \, \omega_1^{\prime}\,t}}
    {(\omega_1^{\prime}-\omega_2^{\prime})^2 \,  (\omega_3^{\prime}-\omega_1^{\prime})}
    \notag\\
    &\quad{}
    -\frac{\ii\,t\,a_1\,a_2\,a_3\,\epsilon^3 \,\e^{-\ii \omega_1^{\prime} t}}
    {(\omega_1^{\prime}-\omega_2^{\prime}) \, (\omega_3^{\prime}-\omega_1^{\prime})}
    -\frac{a_1\,a_2\,a_3\,\epsilon^3 \,\e^{-\ii \omega_2^{\prime} t}}
    {(\omega_1^{\prime}-\omega_2^{\prime})^2 \,  (\omega_2^{\prime}-\omega_3^{\prime})}
    \notag\\
    &\quad{}
    +\frac{a_1\,a_2\,a_3\,\epsilon^3 \,\e^{-\ii \omega_3^{\prime}t}}
    {(\omega_3^{\prime}-\omega_1^{\prime})^2 \, (\omega_2^{\prime}-\omega_3^{\prime})}.
    \label{third_order}
\end{align}

Next, the sixth-order perturbative calculation is also organized as a quantity divided by $\psi_1(0)$ to simplify the expression, as follows: 
\begin{align}
&\frac{\psi_1^{(6)}(t)}{\psi_1(0)}
  \notag\\
  &= T_{\alpha}(2,0,0) + T_{\alpha}(2,1,0) + T_{\alpha}(2,2,0) + T_{\alpha}(2,0,1)\notag\\
&{}+ T_{\alpha}(2,1,1) + T_{\alpha}(2,0,2) + T_{\beta}(2,0,0) + T_{\beta}(2,1,0)\notag\\
&{}+ T_{\beta}(2,0,1) + T_{\gamma}(2,0,0) + T_{\gamma}(2,1,0) + T_{\gamma}(2,0,1)\notag\\
&= \frac{3\,a_1^2 \, a_2^2 \,  a_3^2\,\epsilon^6\,\e^{-\ii \omega _1^{\prime} t}}{(\omega _1^{\prime}-\omega _2^{\prime})^2 \,  (\omega _3^{\prime}-\omega _1^{\prime}){}^4}
-\frac{4\,a_1^2 \, a_2^2 \,  a_3^2 \, \epsilon^6 \, \e^{-\ii \omega _1^{\prime} t}}{(\omega _1^{\prime}-\omega _2^{\prime})^3 \, (\omega _3^{\prime}-\omega _1^{\prime})^3}
\notag\\
&{}
+\frac{3\,a_1^2 \, a_2^2 \,  a_3^2 \, \epsilon^6 \, \e^{-\ii \omega _1^{\prime} t}}{(\omega _1^{\prime}-\omega _2^{\prime})^4 \,  (\omega _3^{\prime}-\omega _1^{\prime})^2}
-\frac{2\,a_1^2 \, a_2^2 \, a_3^2 \, \epsilon^6 \, \ii t \, \e^{-\ii \omega _1^{\prime} t}}{(\omega _1^{\prime}-\omega _2^{\prime})^2 \, (\omega _3^{\prime}-\omega _1^{\prime})^3}
\notag\\
&{}
+\frac{2\,a_1^2 \, a_2^2 \,  a_3^2 \, \epsilon^6 \, \ii t \, \e^{-\ii \omega _1^{\prime} t}}{(\omega _1^{\prime}-\omega _2^{\prime})^3 (\omega _3^{\prime}-\omega _1^{\prime})^2}
+\frac{a_1^2 \, a_2^2 \, a_3^2 \, \epsilon^6 \, (\ii t)^2 \, \e^{-\ii \omega _1^{\prime} t}}{2 \, (\omega _1^{\prime}-\omega _2^{\prime})^2 (\omega _3^{\prime}-\omega _1^{\prime})^2}
\notag\\
&{}
-\frac{3\,a_1^2 \, a_2^2 \,  a_3^2\,\epsilon^6 \, \e^{-\ii \omega _3^{\prime} t}}{(\omega _2^{\prime}-\omega _3^{\prime})^2 \, (\omega _3^{\prime}-\omega _1^{\prime})^4}
+\frac{2\,a_1^2 \, a_2^2 \,  a_3^2 \, \epsilon^6 \, \e^{-\ii \omega _3^{\prime} t}}{(\omega _2^{\prime}-\omega _3^{\prime})^3 \, (\omega _3^{\prime}-\omega _1^{\prime}){}^3}
\notag\\
&{}
-\frac{a_1^2 \, a_2^2 \, a_3^2 \, \epsilon^6 \, \ii t \, \e^{-\ii \omega _3^{\prime} t}}{(\omega _2^{\prime}-\omega _3^{\prime})^2 \, (\omega _3^{\prime}-\omega _1^{\prime})^3}
+\frac{2 \, a_1^2 \, a_2^2 \,  a_3^2 \, \epsilon^6 \, \e^{-\ii \omega _2^{\prime} t}}{(\omega _1^{\prime}-\omega _2^{\prime})^3 \, (\omega _2^{\prime}-\omega _3^{\prime})^3}
\notag\\
&{}
-\frac{3\,a_1^2 \, a_2^2 \,  a_3^2 \, \epsilon^6 \, \e^{-\ii \omega _2^{\prime} t}}{(\omega _1^{\prime}-\omega _2^{\prime})^4 \, (\omega _2^{\prime}-\omega _3^{\prime})^2}
+\frac{a_1^2 \, a_2^2 \, a_3^2 \, \epsilon^6 \, \ii t \, \e^{-\ii \omega _2^{\prime} t}}{(\omega _1^{\prime}-\omega _2^{\prime})^3 \,  (\omega _2^{\prime}-\omega _3^{\prime})^2}.
\label{sixth_order}
\end{align}

Next, the ninth-order perturbative calculation is also organized as a quantity divided by $\psi_1(0)$ to simplify the expression, as follows: 
\begin{align}
&\frac{\psi_1^{(9)}(t)}{\psi_1(0)}
  \notag\\
  &=  T_{\alpha}(3,0,0) + T_{\alpha}(3,1,0) + T_{\alpha}(3,2,0) + T_{\alpha}(3,3,0)\notag\\
&+ T_{\alpha}(3,0,1) + T_{\alpha}(3,1,1) + T_{\alpha}(3,2,1) + T_{\alpha}(3,0,2)\notag\\
&+ T_{\alpha}(3,1,2) + T_{\alpha}(3,0,3) + T_{\beta}(3,0,0) + T_{\beta}(3,1,0)\notag\\
&+ T_{\beta}(3,2,0) + T_{\beta}(3,0,1) + T_{\beta}(3,1,1) + T_{\beta}(3,0,2)\notag\\
&+ T_{\gamma}(3,0,0) + T_{\gamma}(3,1,0) + T_{\gamma}(3,2,0) + T_{\gamma}(3,0,1)\notag\\
&+ T_{\gamma}(3,1,1) + T_{\gamma}(3,0,2)\notag\\
&
= \frac{10\,a_1^3 \, a_2^3 \,  a_3^3 \, \epsilon^9 \,\e^{-\ii \omega _1^{\prime} t}}{(\omega _1^{\prime}-\omega _2^{\prime})^3 \, (\omega _3^{\prime}-\omega _1^{\prime})^6}
-
\frac{18 \, a_1^3 \, a_2^3 \,  a_3^3 \, \epsilon^9 \, \e^{-\ii \omega _1^{\prime} t}}{(\omega _1^{\prime}-\omega _2^{\prime})^4 \, (\omega _3^{\prime}-\omega _1^{\prime})^5}
\notag\\
&{}+
\frac{18 \, a_1^3 \, a_2^3 \,  a_3^3 \, \epsilon^9 \, \e^{-\ii \omega _1^{\prime} t}}{(\omega _1^{\prime}-\omega _2^{\prime})^5 \,  (\omega _3^{\prime}-\omega _1^{\prime})^4}
-
\frac{10 \, a_1^3 \, a_2^3 \,  a_3^3 \, \epsilon^9 \, \e^{-\ii \omega _1^{\prime} t}}{(\omega _1^{\prime}-\omega _2^{\prime})^6 \,  (\omega _3^{\prime}-\omega _1^{\prime}){}^3}
\notag\\
&{}-
\frac{6 \, a_1^3 \, a_2^3 \,  a_3^3 \, \epsilon^9 \, \ii t \, \e^{-\ii \omega _1^{\prime} t}}{(\omega _1^{\prime}-\omega _2^{\prime})^3 \,  (\omega _3^{\prime}-\omega _1^{\prime})^5}
+
\frac{9 \, a_1^3 \, a_2^3 \,  a_3^3 \, \epsilon^9 \, \ii t \, \e^{-\ii \omega _1^{\prime} t}}{(\omega _1^{\prime}-\omega _2^{\prime})^4 \,  (\omega _3^{\prime}-\omega _1^{\prime})^4}
\notag\\
&{}-
\frac{6\,a_1^3 \, a_2^3 \, a_3^3 \, \epsilon^9 \, \ii t \, \e^{-\ii \omega _1^{\prime} t}}{(\omega _1^{\prime}-\omega _2^{\prime})^5 \, (\omega _3^{\prime}-\omega _1^{\prime})^3}
+
\frac{3\,a_1^3 \, a_2^3 \,  a_3^3 \, \epsilon^9 \, (\ii t)^2 \, \e^{-\ii \omega _1^{\prime} t}}{2 \,(\omega _1^{\prime}-\omega _2^{\prime})^3 \, (\omega _3^{\prime}-\omega _1^{\prime})^4}
\notag\\
&{}
-
\frac{3\,a_1^3 \, a_2^3 \, a_3^3 \, \epsilon^9 \, (\ii t)^2 \, \e^{-\ii \omega _1^{\prime} t}}{2 \,(\omega _1^{\prime}-\omega _2^{\prime})^4 \, (\omega _3^{\prime}-\omega _1^{\prime})^3}
-
\frac{a_1^3 \, a_2^3 \, a_3^3 \, \epsilon^9 \, (\ii t)^3 \, \e^{-\ii \omega _1^{\prime} t}}{6 \, (\omega _1^{\prime}-\omega _2^{\prime})^3 \, (\omega _3^{\prime}-\omega _1^{\prime})^3}
\notag\\
&{}+
\frac{10 \, a_1^3 \, a_2^3 \,  a_3^3 \, \epsilon^9 \, \e^{-\ii \omega _3^{\prime} t}}{(\omega _2^{\prime}-\omega _3^{\prime})^3 \, (\omega _3^{\prime}-\omega _1^{\prime})^6}
-
\frac{12 \, a_1^3 \, a_2^3 \, a_3^3 \, \epsilon^9 \, \e^{-\ii \omega _3^{\prime} t}}{(\omega _2^{\prime}-\omega _3^{\prime})^4 \, (\omega _3^{\prime}-\omega _1^{\prime})^5}
\notag\\
&{}+
\frac{6 \, a_1^3 \, a_2^3 \, a_3^3 \, \epsilon^9 \, \e^{-\ii \omega _3^{\prime} t}}{(\omega _2^{\prime}-\omega _3^{\prime})^5 \, (\omega _3^{\prime}-\omega _1^{\prime})^4}
+
\frac{4 \, a_1^3 \, a_2^3 \,  a_3^3 \, \epsilon^9 \, \ii t \, \e^{-\ii \omega _3^{\prime} t}}{(\omega _2^{\prime}-\omega _3^{\prime})^3 \, (\omega _3^{\prime}-\omega _1^{\prime})^5}
\notag\\
&{}-
\frac{3 \, a_1^3 \, a_2^3 \, a_3^3 \, \epsilon^9 \, \ii t \, \e^{-\ii \omega _3^{\prime} t}}{(\omega _2^{\prime}-\omega _3^{\prime})^4 \, (\omega _3^{\prime}-\omega _1^{\prime})^4}
+
\frac{a_1^3 \, a_2^3 \, a_3^3 \, \epsilon^9 \, (\ii t)^2 \, \e^{-\ii \omega _3^{\prime} t}}{2 \, (\omega _2^{\prime}-\omega _3^{\prime})^3 \, (\omega _3^{\prime}-\omega _1^{\prime})^4}
\notag\\
&{}-
\frac{6 \, a_1^3 \, a_2^3 \, a_3^3 \, \epsilon^9 \, \e^{-\ii \omega _2^{\prime} t}}{(\omega _1^{\prime}-\omega _2^{\prime})^4 \, (\omega _2^{\prime}-\omega _3^{\prime})^5}
+
\frac{12 \, a_1^3 \, a_2^3 \, a_3^3 \, \epsilon^9 \, \e^{-\ii \omega _2^{\prime} t}}{(\omega _1^{\prime}-\omega _2^{\prime})^5 \,  (\omega _2^{\prime}-\omega _3^{\prime})^4}
\notag\\
&{}-
\frac{10 \, a_1^3 \, a_2^3 \,  a_3^3 \, \epsilon^9 \, \e^{-\ii \omega _2^{\prime} t}}{(\omega _1^{\prime}-\omega _2^{\prime})^6 \,  (\omega _2^{\prime}-\omega _3^{\prime})^3}
-
\frac{3 \, a_1^3 \, a_2^3 \, a_3^3 \, \epsilon^9 \, \ii t \, \e^{-\ii \omega _2^{\prime} t}}{(\omega _1^{\prime}-\omega _2^{\prime})^4 \,  (\omega _2^{\prime}-\omega _3^{\prime})^4}
\notag\\
&{}+
\frac{4 \, a_1^3 \, a_2^3 \,  a_3^3 \, \epsilon^9 \, \ii t \, \e^{-\ii \omega _2^{\prime} t}}{(\omega _1^{\prime}-\omega _2^{\prime})^5 \, (\omega _2^{\prime}-\omega _3^{\prime})^3}
-
\frac{a_1^3 \, a_2^3 \, a_3^3 \, \epsilon^9 \, (\ii t)^2 \, \e^{-\ii \omega _2^{\prime} t}}{2 \, (\omega _1^{\prime}-\omega _2^{\prime})^4 \, (\omega _2^{\prime}-\omega _3^{\prime})^3}.
\label{ninth_order}
\end{align}

Next, we describe regularities that can be found in the results of the perturbative calculations. 
We infer the general form of the perturbative expansion by adding up the appropriate terms in the organized expressions. 
The zero-order perturbation solution is represented by a single term as follows: 
\[
\psi_1^{(0)}(t) = T_{\alpha}(0,0,0) \, \psi_1(0) = \e^{-\ii \omega_1^{\prime} t}\, \psi_1(0). 
\]
Based on the above, consider this sum: 
\begin{align}
    &\sum_{m=0}^\infty T_{\alpha}(m, 0, m) 
    \notag\\
    &=T_{\alpha}(0, 0, 0) + T_{\alpha}(1, 0, 1) + T_{\alpha}(2, 0, 2) + \cdots
    \notag\\
    &=\e^{-\ii \omega_1^{\prime} t} \Bigg[ 1  
    -\frac{a_1\,a_2\,a_3\,\epsilon^3\,\ii t}{(\omega_1^{\prime}-\omega_2^{\prime})(\omega_3^{\prime}-\omega_1^{\prime})}
  \notag\\
  &\qquad\qquad{}
  +\frac{1}{2!}\,\frac{a_1^2\,a_2^2\,a_3^2\,\epsilon^6\,(\ii t)^2}{(\omega_1^{\prime}-\omega_2^{\prime})^2(\omega_3^{\prime}-\omega_1^{\prime})^2}
  \notag\\
  &\qquad\qquad{}
  -\frac{1}{3!}\,\frac{a_1^3\,a_2^3\,a_3^3\,\epsilon^{9}\,(\ii t)^3}{(\omega_1^{\prime}-\omega_2^{\prime})^3(\omega_3^{\prime}-\omega_1^{\prime})^3}
+\cdots \Bigg]. 
\end{align}
Now, we define 
\begin{align}
X := \frac{a_1\,a_2\,a_3\,\epsilon^3}{(\omega_1^{\prime}-\omega_2^{\prime})(\omega_3^{\prime}-\omega_1^{\prime})}, 
\label{X}
\end{align}
so 
\begin{align}
    &\sum_{m=0}^\infty T_{\alpha}(m, 0, m) 
    \notag\\
    &=\e^{-\ii \omega_1^{\prime} t} 
    \left(1 - (\ii \, X \, t) + \frac{1}{2!}\,(\ii \, X \, t)^2 - \frac{1}{3!}\,(\ii \, X \, t)^3 +\cdots\right) \notag\\
    &= \e^{-\ii \omega_1^{\prime} t} \, \e^{-\ii \,X\, t}. 
  \label{k0l0}
\end{align}
Next, the summation with $T_{\alpha}(1,0,0)$ as the initial term is as follows: 
\begin{align}
&\sum_{m=0}^\infty T_{\alpha}(1+m, 0, m) 
    \notag\\
  &=T_{\alpha}(1, 0, 0) + T_{\alpha}(2, 0, 1) + T_{\alpha}(3, 0, 2) + \cdots
  \notag\\
  &=\frac{ a_1\,a_2\,a_3\,\epsilon^3\,\e^{-\ii \omega_1^{\prime} t}}{(\omega_1^{\prime}-\omega_2^{\prime})\, (\omega_3^{\prime}-\omega_1^{\prime})^2}
  -\frac{2\,a_1^2\,a_2^2\,a_3^2\,\epsilon^6\,\ii t \e^{-\ii\omega_1^{\prime} t}}{(\omega_1^{\prime}-\omega_2^{\prime})^2\,(\omega_3^{\prime}-\omega_1^{\prime})^3}
  \notag\\
  &
  +\frac{3\,a_1^3\,a_2^3\,a_3^3\,\epsilon^9\,(\ii t)^2\,\e^{-\ii\omega_1^{\prime} t}}{2\,(\omega_1^{\prime}-\omega_2^{\prime})^3\,(\omega_3^{\prime}-\omega_1^{\prime})^4}
  -\frac{4\,a_1^4\,a_2^4\,a_3^4\,\epsilon^{12}\,(\ii t)^3 \, \e^{-\ii \omega_1^{\prime} t}}{6\,(\omega_1^{\prime}-\omega_2^{\prime})^4\,(\omega_3^{\prime}-\omega_1^{\prime})^5}
  \notag\\
  &+\cdots
  \notag\\
  &= \frac{\e^{-\ii \omega_1^{\prime} t} \, X}{(\omega_3^{\prime}-\omega_1^{\prime})} \,\Big(1-2\,(\ii\,X\,t) +\frac{3}{2!}\,(\ii\,X\,t)^2 \notag\\
  &\qquad\qquad\qquad\qquad\qquad\quad {}- \frac{4}{3!}\,(\ii\,X\,t)^3 + \cdots \Big). 
  \label{k1l0}
\end{align}
Next, the summation with $T_{\alpha}(2,0,0)$ as the initial term is as follows: 
\begin{align}
&\sum_{m=0}^\infty T_{\alpha}(2+m, 0, m) 
    \notag\\
  &=T_{\alpha}(2, 0, 0) + T_{\alpha}(3, 0, 1) + T_{\alpha}(4, 0, 2) + \cdots\notag\\
  &
  \frac{3\,a_1^2 \, a_2^2 \,  a_3^2\,\epsilon^3\,\exp(-\ii \omega _1^{\prime} t)}{(\omega _1^{\prime}-\omega _2^{\prime})^2 \, (\omega _3^{\prime}-\omega _1^{\prime})^4}
  -\frac{6\,a_1^3 \, a_2^3 \,  a_3^3\,\epsilon^9\, \ii t \,\exp(-\ii \omega _1^{\prime} t)}{(\omega _1^{\prime}-\omega _2^{\prime})^3 \, (\omega _3^{\prime}-\omega _1^{\prime})^5}\notag\\
  &
  +\frac{10\,a_1^4 \, a_2^4 \, a_3^4 \, \epsilon^{12} \, (\ii t)^2 \, \exp(-\ii \omega _1^{\prime} t)}{2\,(\omega _1^{\prime}-\omega _2^{\prime})^4 \, (\omega _3^{\prime}-\omega _1^{\prime})^6}
  +\cdots\notag\\
  &= \frac{\e^{-\ii \omega_1^{\prime} t} \, X^2}{(\omega_3^{\prime}-\omega_1^{\prime})^2} \,\Big(3-6\,(\ii\,X\,t) +\frac{10}{2!}\,(\ii\,X\,t)^2  + \cdots \Big). 
  \label{k2l0}
\end{align}
By deriving the regularity from equations \eqref{k1l0} and \eqref{k2l0}, for the summation with $T_{\alpha}(k,0,0)$ $(k\not=0)$ as the initial term, we obtain 
\begin{align}
&\sum_{m=0}^{\infty} T_{\alpha}(k+m,0,m)
\notag\\
    &=\frac{\e^{-\ii \omega_1^{\prime} t} \, X^k}{(\omega_3^{\prime}-\omega_1^{\prime})^k} \,\binom{2k-1}{k}\,
    \sum_{m=0}^{\infty}
    \frac{(2k)_m}{(k)_m}
    \frac{1}{m!}
    (-\ii\,X\,t)^m
    \notag\\
    &=
    \frac{\e^{-\ii \omega_1^{\prime} t} \, X^k}{(\omega_3^{\prime}-\omega_1^{\prime})^k} \,\binom{2k-1}{k}\,
    {}_1F_1\left(2 k;k;-\ii\,X\,t\right), 
    \label{l0_deformation}
\end{align}
where $(x)_{m}$ is the Pochhammer symbol defined as 
\begin{align*}
    (x)_0 := 1,\quad (x)_m := \prod_{y=0}^{m-1}(x+y)\quad(m=1,\,2,\,\dots). 
\end{align*}
Also, the transformation from the first line to the second line uses the hypergeometric series as follows: 
\begin{align*}
    {}_1F_1(a;b;z):=\sum_{\ell = 0}^{\infty}\frac{(a)_{\ell}}{(b)_{\ell}}\frac{z^{\ell}}{\ell !}. 
\end{align*}

By deriving the regularity for $l=1$ by the same procedure as from \eqref{k0l0} to \eqref{l0_deformation}, we obtain 
\begin{align}
 &\sum_{m=0}^{\infty} T_{\alpha}(k+m,1,m)
\notag\\
   &=-\frac{\e^{-\ii \omega_1^{\prime} t} \, X^k}{(\omega_3^{\prime}-\omega_1^{\prime})^k} \,\frac{\omega_3^{\prime}-\omega_1^{\prime}}{\omega_1^{\prime}-\omega_2^{\prime}}\,
    \binom{2k-2}{k-1}\,
    \binom{k}{1}
    \notag\\
    &\times
    \sum_{m=0}^{\infty}
    \frac{(2k-1)_m\,(k+1)_m}{(k)_m\,(k)_m}\,
    \frac{1}{m!}\,
    \left(-\ii\,X\,t\right)^m
    \notag\\
    &=
    -\frac{\e^{-\ii \omega_1^{\prime} t} \, X^k}{(\omega_3^{\prime}-\omega_1^{\prime})^k}\,\frac{\omega_3^{\prime}-\omega_1^{\prime}}{\omega_1^{\prime}-\omega_2^{\prime}}\,
    \binom{2k-2}{k-1}\,
    \binom{k}{1}\,
    \notag\\
    &\times
    {}_2F_2(2k-1, k+1 ; k, k ; -\ii\,X\,t), 
    \label{l1_deformation}
\end{align}
where the transformation from the first line to the second line uses the hypergeometric series, 
\begin{align}
    {}_2F_2(a, b ; c, d ; z):=\sum_{\ell = 0}^{\infty}\frac{(a)_{\ell}\,(b)_{\ell}}{(c)_{\ell}\,(d)_{\ell}}\frac{z^{\ell}}{\ell !}. 
\end{align}
Similarly, the case of $l=2$ is as follows: 
\begin{align}
&\sum_{m=0}^{\infty} T_{\alpha}(k+m,2,m)
\notag\\
   &=\frac{\e^{-\ii \omega_1^{\prime} t} \, X^k}{(\omega_3^{\prime}-\omega_1^{\prime})^k} \left(\frac{\omega_3^{\prime}-\omega_1^{\prime}}{\omega_1^{\prime}-\omega_2^{\prime}}\right)^2
    \binom{2k-3}{k-2}\,
    \binom{k+1}{2}\,
    \notag\\
    &\times
    \sum_{m=0}^{\infty}
    \frac{(2k-2)_m\,(k+2)_m}{(k)_m\,(k)_m}\,
    \frac{1}{m!}\,
    (-\ii\,X\,t)^m
    \notag\\
    &=
    \frac{\e^{-\ii \omega_1^{\prime} t} \, X^k}{(\omega_3^{\prime}-\omega_1^{\prime})^k} \left(\frac{\omega_3^{\prime}-\omega_1^{\prime}}{\omega_1^{\prime}-\omega_2^{\prime}}\right)^2
    \binom{2k-3}{k-2}\,
    \binom{k+1}{2}\,
    \notag\\
    &\times
    {}_2F_2(2k-2, k+2 ; k, k ; -\ii\,X\,t). 
    \label{l2_deformation}
\end{align}
From expressions \eqref{l0_deformation}, \eqref{l1_deformation}, and \eqref{l2_deformation}, except for the case where $k=l=0$ (the series of \eqref{k0l0}), we can derive the following regularity for general $k$, $l$ $(k,\,l=0,\,1,\,\dots)$, 
\begin{align}
&\sum_{m=0}^{\infty} T_{\alpha}(k+m,l,m)
\notag\\
   &=(-1)^l\,\frac{\e^{-\ii \omega_1^{\prime} t} \, X^k}{(\omega_3^{\prime}-\omega_1^{\prime})^k} \left(\frac{\omega_3^{\prime}-\omega_1^{\prime}}{\omega_1^{\prime}-\omega_2^{\prime}}\right)^l
    \notag\\
    &\quad\times
    \binom{2k-l-1}{k-l}\,
    \binom{k+l-1}{l}\,
    \notag\\
    &\quad\times
    {}_2F_2(2k-l, k+l ; k, k ; -\ii\,X\,t). 
    \label{sum_T(k+m,l,m)}
\end{align}
When $k=l=0$, the right-hand side of \eqref{sum_T(k+m,l,m)} is $\e^{-\ii \omega_1^{\prime} t}$. 
Considering this, the sum of the perturbations on $k$ for $n=3k$ up to infinite order is as follows: 
\begin{align}
&\sum_{k=0}^{\infty}\sum_{m=0}^{k}\sum_{l=0}^{k-m} T_{\alpha}(k,l,m) 
=\sum_{k=0}^{\infty}\sum_{l=0}^{k}\sum_{m=0}^{\infty} T_{\alpha}(k+m,l,m)
\notag\\
  &=\sum _{k=0}^{\infty }
  \Biggl[
  \sum _{l=0}^{k} 
  (-1)^l\,\frac{\e^{-\ii \omega_1^{\prime} t} \, X^k}{(\omega_3^{\prime}-\omega_1^{\prime})^k} \left(\frac{\omega_3^{\prime}-\omega_1^{\prime}}{\omega_1^{\prime}-\omega_2^{\prime}}\right)^l
    \notag\\
    &\quad\times
    \binom{2k-l-1}{k-l}\,
    \binom{k+l-1}{l}\,
    \notag\\
    &\quad\times
    {}_2F_2(2k-l, k+l ; k, k ; -\ii\,X\,t)
\Biggr]
  \notag
  \\
  &+\e^{-\ii \omega_1^{\prime} t}(\e^{-\ii \, X \, t}-1),  
  \label{sum_T(k,l,m)}
\end{align}
where binomial coefficients taking negative arguments follow the definition in reference~\cite{kronenburg2011binomial}.
So far, we have only added up the terms whose order of perturbative expansion is $n = 3k$ $(k=0,\,\,1,\,\dots)$ and that are proportional to $\e^{-\ii\omega_1^{\prime} t}$. 
There are also terms proportional to $\e^{-\ii \omega_3^{\prime} t}$ and $\e^{-\ii \omega_2^{\prime} t}$, and considering the case where the order of the perturbative expansion is $n = 3k+1$ and $n = 3k+2$, to calculate $\psi_ 1(t)$ of \eqref{psi_1}, we need to use nine expressions in total, including the expression \eqref{sum_T(k,l,m)}. 
For the other eight expressions, the same argument can be used to find the regularity as in expressions \eqref{deformation_step3} to \eqref{sum_T(k,l,m)}. 
Let expression \eqref{sum_T(k,l,m)} other than $\e^{-\ii \omega_1^{\prime} t}$ be $A_1$, which is as follows: 
\[
\sum_{k=0}^{\infty}\sum_{m=0}^{k}\sum_{l=0}^{k-m} T_{\alpha}(k,l,m) 
= A_1\,\psi_1(0)\,\e^{-\ii \omega_1^{\prime} t}.
\]
Similarly, the terms for $\e^{-\ii \omega_1^{\prime} t}$, $\e^{-\ii \omega_3^{\prime} t}$, and $\e^{-\ii \omega_2^{\prime} t}$ are $A_\mu$, $B_\mu$, and $C_\mu$ $(\mu =1,\,\,2,\,3)$, respectively; terms with perturbation orders of $n=3k$, $n=3k+1$, and $n=3k+2$ are denoted by the subscript $\mu$ as $1$, $3$, and $2$, respectively. 
It follows that the overall structure of the solution $\psi_1(t)$ for state $1$ can be written as follows: 
\begin{align}
  \psi_1(t)&=
  \left(A_1\,\psi_1(0)+A_3\,\psi_3(0)+A_2\,\psi_2(0)\right)\,\e^{-\ii \omega^{\prime}_1 t}\notag\\
  &\quad+\left(B_1\,\psi_1(0)+B_3\,\psi_3(0)+B_2\,\psi_2(0)\right)\,\e^{-\ii \omega^{\prime}_3 t}\notag\\
  &\quad+\left(C_1\,\psi_1(0)+C_3\,\psi_3(0)+C_2\,\psi_2(0)\right)\,\e^{-\ii \omega^{\prime}_2 t}. 
  \label{form}
\end{align}
In the following, we describe $A_1$, $A_3$, $A_2$, $B_1$, $B_3$, $B_2$, $C_1$, $C_3$, and $C_2$ in detail. 
As in \eqref{X}, we define $Y$ and $Z$ as follows: 
\begin{align*}
  Y&:=\frac{a_1 \, a_2 \, a_3\,\epsilon^3 }{(\omega_2^{\prime}-\omega_3^{\prime})\, (\omega_3^{\prime}-\omega_1^{\prime})}\\
  Z&:=\frac{a_1 \, a_2 \, a_3\,\epsilon^3}{(\omega_1^{\prime}-\omega_2^{\prime})\, (\omega_2^{\prime}-\omega_3^{\prime})}. 
\end{align*}
Then, each expression can be written as follows: 
\begin{align}
A_1&=\notag\\
  &\sum _{k=0}^{\infty }
  \Biggl[
  \sum _{l=0}^{k} 
  (-1)^l\,\frac{X^k}{(\omega_3^{\prime}-\omega_1^{\prime})^k} \left(\frac{\omega_3^{\prime}-\omega_1^{\prime}}{\omega_1^{\prime}-\omega_2^{\prime}}\right)^l
    \notag\\
    &\quad\times
    \binom{2k-l-1}{k-l}\,
    \binom{k+l-1}{l}\,
    \notag\\
    &\quad\times
    {}_2F_2(2k-l, k+l ; k, k ; -\ii\,X\,t)
\Biggr]
  \notag
  \\
  &+\e^{-\ii \, X \, t}-1, 
  \label{a1}\\
A_3&=\notag\\
  &\sum _{k=0}^{\infty }
  \Biggl[
  \sum _{l=0}^{k} 
  (-1)^l\,
  \frac{a_3\,\epsilon}{\omega_3'-\omega_1'}
  \frac{X^k}{(\omega_3'-\omega_1')^k} \left(\frac{\omega_3'-\omega_1'}{\omega_1'-\omega_2'}\right)^l
    \notag\\
    &\quad\times
    \binom{2k-l}{k-l}\,
    \binom{k+l-1}{l}\,
    \notag\\
    &\quad\times
    {}_2F_2(2k-l+1, k+l ; k, k+1 ; -\ii\,X\,t)
\Biggr], 
\label{a3}\\
A_2&=\notag\\
  &\sum_{k=0}^{\infty }
  \Biggl[
  \sum _{l=0}^{k} 
  (-1)^{l+1}\,
  \frac{a_2\,a_3\,\epsilon^2}{(\omega_1'-\omega_2')(\omega_3'-\omega_1')}
  \frac{X^k}{(\omega_3'-\omega_1')^k}
  \notag\\
  &\quad\times\left(\frac{\omega_3'-\omega_1'}{\omega_1'-\omega_2'}\right)^l
    \binom{2k-l}{k-l}\,
    \binom{k+l}{l}\,
    \notag\\
    &\quad\times
    {}_2F_2(2k-l+1, k+l+1 ; k+1, k+1 ; -\ii\,X\,t)
\Biggr], 
\label{a2}\\
B_1&=\notag\\
  &\sum _{k=1}^{\infty }
  \Biggl[
  \sum _{l=0}^{k-1} 
  (-1)^{k+l+1}\,
  \frac{Y^k}{(\omega_3'-\omega_1')^k} \left(\frac{\omega_3'-\omega_1'}{\omega_2'-\omega_3'}\right)^l
    \notag\\
    &\quad\times
    \binom{2k-l-1}{k-l-1}\,
    \binom{k+l-1}{l}\,
    \notag\\
    &\quad\times
    {}_2F_2(2k-l, k+l ; k, k+1 ; -\ii\,Y\,t)
\Biggr], 
\label{b1}\\
B_3&=\notag\\
  &\sum _{k=0}^{\infty }
  \Biggl[
  \sum _{l=0}^{k} 
  (-1)^{k+l+1}\,
  \frac{a_3\,\epsilon}{\omega_3'-\omega_1'}
  \frac{Y^k}{(\omega_3'-\omega_1')^k} \left(\frac{\omega_3'-\omega_1'}{\omega_2'-\omega_3'}\right)^l
    \notag\\
    &\quad\times
    \binom{2k-l}{k-l}\,
    \binom{k+l-1}{l}\,
    \notag\\
    &\quad\times
    {}_2F_2(2k-l+1, k+l ; k, k+1 ; -\ii\,Y\,t)
\Biggr], 
\label{b3}\\
B_2&=\notag\\
  &\sum_{k=0}^{\infty }
  \Biggl[
  \sum _{l=0}^{k} 
  (-1)^{k+l+1}\,
  \frac{a_2\,a_3\,\epsilon^2}{(\omega_2'-\omega_3')(\omega_3'-\omega_1')}
  \frac{Y^k}{(\omega_3'-\omega_1')^k}
  \notag\\
  &\quad\times\left(\frac{\omega_3'-\omega_1'}{\omega_2'-\omega_3'}\right)^l
    \binom{2k-l}{k-l}\,
    \binom{k+l}{l}\,
    \notag\\
    &\quad\times
    {}_2F_2(2k-l+1, k+l+1 ; k+1, k+1 ; -\ii\,Y\,t)
\Biggr], 
\label{b2}\\
C_1&=\notag\\
  &\sum _{k=1}^{\infty }
  \Biggl[
  \sum _{l=0}^{k-1} 
  (-1)^{l+1}\,
  \frac{Z^k}{(\omega_1'-\omega_2')^k} \left(\frac{\omega_1'-\omega_2'}{\omega_2'-\omega_3'}\right)^l
    \notag\\
    &\quad\times
    \binom{2k-l-1}{k-l-1}\,
    \binom{k+l-1}{l}\,
    \notag\\
    &\quad\times
    {}_2F_2(2k-l, k+l ; k, k+1 ; -\ii\,Z\,t)
\Biggr], 
\label{c1}\\
C_3&=\notag\\
  &\sum _{k=1}^{\infty }
  \Biggl[
  \sum _{l=0}^{k-1} 
  (-1)^{l}\,
  \frac{a_3\,\epsilon}{\omega_2'-\omega_3'}
  \frac{Z^k}{(\omega_1'-\omega_2')^k} \left(\frac{\omega_1'-\omega_2'}{\omega_2'-\omega_3'}\right)^l
    \notag\\
    &\quad\times
    \binom{2k-l-1}{k-l-1}\,
    \binom{k+l}{l}\,
    \notag\\
    &\quad\times
    {}_2F_2(2k-l, k+l+1 ; k+1, k+1 ; -\ii\,Z\,t)
\Biggr], 
\label{c3}\\
C_2&=\notag\\
  &\sum_{k=0}^{\infty }
  \Biggl[
  \sum _{l=0}^{k} 
  (-1)^{k+l+1}\,
  \frac{a_2\,a_3\,\epsilon^2}{(\omega_1'-\omega_2')(\omega_2'-\omega_3')}
  \frac{Z^k}{(\omega_1'-\omega_2')^k}
  \notag\\
  &\quad\times
  \left(\frac{\omega_1'-\omega_2'}{\omega_2'-\omega_3'}\right)^l
    \binom{2k-l}{k-l}\,
    \binom{k+l}{l}\,
    \notag\\
    &\quad\times
    {}_2F_2(2k-l+1, k+l+1 ; k+1, k+1 ; -\ii\,Z\,t)
\Biggr]. 
\label{c2}
\end{align}

\subsection{Explicit Expression of Eigenfrequency and Higher-order Correction}
\label{highorder}
In the previous section, we formulated perturbative expansions up to infinite order using hypergeometric series, such as expressions \eqref{a1} to \eqref{c2}.
Using these expressions, we then investigate the change in eigenvalues due to the effect of the one-way link graph.
Since the square root of the eigenvalue of the Laplacian matrix is eigenfrequency, we discuss eigenfrequency hereafter. 

For example, the change in eigenfrequency $\omega_1$ can be obtained from the change in the coefficient of the derivative of $\psi_1(t)$ with time, so if the first term on the right-hand side of \eqref{form} is organized as follows: 
\[
\left(A_1\,\psi_1(0)+A_3\,\psi_3(0)+A_2\,\psi_2(0)\right)\,\e^{-\ii \omega^{\prime}_1 t} \propto \e^{-\ii \hat{\omega}_1 t}, 
\]
then, $\hat{\omega}_1$ in the exponent is the eigenfrequency after the change. 

With the use of the hypergeometric series, the part corresponding to the eigenfrequency is not explicitly expressed. 
For this reason, we expand it to explicitly show the part corresponding to $\e^{-\ii \hat{\omega}_1 t}$. 

By expanding equation \eqref{a1} and writing it specifically for $k=0,\,1,\,2,\,3$, we get 
\begin{align}
A_1 &= \e^{-\ii X t} \times \textcolor{red}{1} 
\notag\\
&\quad{}+\e^{-\ii X t} \, \frac{X \,(1\textcolor{red}{-\ii X t})}{\omega _3^{\prime}-\omega _1^{\prime}}
\notag\\
&\quad{}\textcolor{red}{-}\e^{-\ii X t} \, \frac{X \, (1\textcolor{red}{-\ii X t})}{\omega _3^{\prime}-\omega _1^{\prime}}\,\frac{\omega_3^{\prime}-\omega_1^{\prime}}{\omega_1^{\prime}-\omega _2^{\prime}}
\notag\\
&\quad{}+\e^{-\ii X t} \, \frac{X^2 \, (6-6\,\ii X t+\textcolor{red}{(-\ii X t)^2})}{2\,(\omega _3^{\prime}-\omega _1^{\prime})^2}
\notag\\
&\quad{}\textcolor{red}{-}\e^{-\ii X t} \, \frac{X^2 \, (4-5\,\ii X t + \textcolor{red}{(-\ii X t)^2})}{(\omega _3^{\prime}-\omega _1^{\prime})^2}\,\frac{\omega_3^{\prime}-\omega_1^{\prime}}{\omega_1^{\prime}-\omega _2^{\prime}}
\notag\\
&\quad{}+\e^{-\ii X t} \, \frac{X^2 \, (6-6\,\ii X t+\textcolor{red}{(-\ii X t)^2})}{2\,(\omega _3^{\prime}-\omega _1^{\prime})^2}\left(\frac{\omega_3^{\prime}-\omega_1^{\prime}}{\omega_1^{\prime}-\omega _2^{\prime}}\right)^2
\notag\\
&\quad{}+\e^{-\ii X t} \, \frac{X^3 \, (60-50\,\ii X t+15\,(-\ii X t)^2+\textcolor{red}{(-\ii X t)^3})}{6\,(\omega _3^{\prime}-\omega _1^{\prime})^3}
\notag\\
&\quad{}\textcolor{red}{-}\e^{-\ii X t} \, \frac{X^3 \, (36-44\,\ii X t+13\,(-\ii X t)^2+\textcolor{red}{(-\ii X t)^3})}{2\,(\omega _3^{\prime}-\omega _1^{\prime})^3}\,
\notag\\
&\qquad\qquad\qquad\qquad\qquad\qquad\qquad
\times\frac{\omega_3^{\prime}-\omega_1^{\prime}}{\omega_1^{\prime}-\omega _2^{\prime}}
\notag\\
&\quad{}+\e^{-\ii X t} \, \frac{X^3 \, (36-44\,\ii X t+13\,(-\ii X t)^2+\textcolor{red}{(-\ii X t)^3})}{2\,(\omega _3^{\prime}-\omega _1^{\prime})^3}
\notag\\
&\qquad\qquad\qquad\qquad\qquad\qquad\qquad
\times\left(\frac{\omega_3^{\prime}-\omega_1^{\prime}}{\omega_1^{\prime}-\omega _2^{\prime}}\right)^2
\notag\\
&\quad{}\textcolor{red}{-}\e^{-\ii X t} \, \frac{X^3 \, (60-50\,\ii X t+15\,(-\ii X t)^2+\textcolor{red}{(-\ii X t)^3})}{6\,(\omega _3^{\prime}-\omega _1^{\prime})^3}
\notag\\
&\qquad\qquad\qquad\qquad\qquad\qquad\qquad
\times\left(\frac{\omega_3^{\prime}-\omega_1^{\prime}}{\omega_1^{\prime}-\omega _2^{\prime}}\right)^3. 
\label{A1k1}
\end{align}
Now, by approximating $A_1$ to the first term on the right-hand side of \eqref{A1k1}, 
\[
A_1 \simeq \e^{-\ii X t}, 
\]
then, 
\begin{align}
A_1 \,\e^{-\ii \omega^{\prime}_1 t} \propto \e^{-\ii (\omega^{\prime}_1 + X)\,t}. 
\label{A_1-1pp0}
\end{align}
Thus the eigenfrequency change $\hat{\omega}_1$ can be inferred as follows: 
\begin{align}
\hat{\omega}_1 \simeq \omega^{\prime}_1 + X. 
\label{app0}
\end{align}
As a higher-order correction to this, by considering the partial sum of the maximum order of each term shown in red on the right-hand side of \eqref{A1k1}, we obtain 
\begin{align}
A_1 &\simeq \e^{-\ii X t} \, \Bigg[1 -\ii\,X^2 \left(\frac{1}{\omega^\prime_3-\omega^\prime_1}-\frac{1}{\omega^\prime_1-\omega^\prime_2}\right)t
\notag\\
&\quad\quad{}
+\frac{1}{2} \left(-\ii\,X^2 \left(\frac{1}{\omega^\prime_3-\omega^\prime_1}-\frac{1}{\omega^\prime_1-\omega^\prime_2}\right)t\right)^2
\notag\\
&\quad\quad{}
+\frac{1}{3!} \left(-\ii\,X^2 \left(\frac{1}{\omega^\prime_3-\omega^\prime_1}-\frac{1}{\omega^\prime_1-\omega^\prime_2}\right)t\right)^3 + \cdots\Bigg]
\notag\\
&= \e^{-\ii X t} \, \sum_{k=0}^\infty \frac{1}{k!}\,\left(-\ii\,X^2 \left(\frac{1}{\omega^\prime_3-\omega^\prime_1}-\frac{1}{\omega^\prime_1-\omega^\prime_2}\right)t\right)^k
\notag\\
&= \e^{-\ii X t} \, \exp\left(-\ii\,X^2 \left(\frac{1}{\omega^\prime_3-\omega^\prime_1}-\frac{1}{\omega^\prime_1-\omega^\prime_2}\right)t\right)
\notag\\
&= \exp\left(-\ii \left(X+ \frac{X^2}{\omega^\prime_3-\omega^\prime_1}-\frac{X^2}{\omega^\prime_1-\omega^\prime_2}\right)t\right). 
\label{A_1-1pp1}
\end{align}
From this result, the eigenfrequency change $\hat{\omega}_1$ can be inferred as follows: 
\begin{align}
\hat{\omega}_1 \simeq \omega^{\prime}_1 + X + \frac{X^2}{\omega_3^{\prime}-\omega _1^{\prime}} -\frac{X^2}{\omega_1^{\prime}-\omega _2^{\prime}}. 
\label{app1}
\end{align}

By considering other series of \eqref{A1k1}, we can expect to be able to improve the precision of the higher-order correction of $A_1$. 
However, since the terms of the first order of $X$ in the second and third terms on the right-hand side of \eqref{A1k1} for example, cannot be taken into account after this higher-order correction, we take them into consideration in this step as follows: 
\[
\e^{-\ii X t} \, \exp\left(\frac{X\,(1-\ii Xt)}{\omega^\prime_3-\omega^\prime_1}-\frac{X\,(1-\ii Xt)}{\omega^\prime_1-\omega^\prime_2}\right). 
\]
With this idea, the higher-order terms are also expressed by exponential functions.
If we write only the relevant part of the exponent of the exponential function, the additional correction terms are as follows: 
\begin{align}
A_1 &\simeq \e^{-\ii X t}
\notag\\
& \times \exp\Bigg(\frac{X(1-\ii X t)}{\omega^\prime_3-\omega^\prime_1}-\frac{X(1-\ii X t)}{\omega^\prime_1-\omega^\prime_2}
+\frac{1}{2}\,\frac{X^2(5-4\,\ii X t)}{(\omega^\prime_3-\omega^\prime_1)^2}
\notag\\
&\quad{}+\frac{-3\,X^2(1-\ii X t)}{(\omega^\prime_3-\omega^\prime_1)\,(\omega^\prime_1-\omega^\prime_2)}
+\frac{1}{2}\,\frac{X^2(5- 4\,\ii X t)}{(\omega^\prime_1-\omega^\prime_2)^2}
\notag\\
&\quad{}+\frac{1}{3}\,\frac{X^3\,(22- 10\,\ii X t)}{(\omega^\prime_3-\omega^\prime_1)^3}
+\frac{-X^3\,(12- 10\,\ii X t)}{(\omega^\prime_3-\omega^\prime_1)^2\,(\omega^\prime_1-\omega^\prime_2)}
\notag\\
&\quad{}+\frac{X^3\,(12- 10\,\ii X t)}{(\omega^\prime_3-\omega^\prime_1)\,(\omega^\prime_1-\omega^\prime_2)^2}
+\frac{1}{3}\,\frac{-X^3\,(22-10\,\ii X t)}{(\omega^\prime_1-\omega^\prime_2)^3}\Bigg). 
\label{A_1-app2}
\end{align}
By extracting the part proportional to $-\ii t$ from the exponential part, the change in eigenfrequency can be inferred as follows: 
\begin{align}
\hat{\omega}_1 &\simeq \omega^{\prime}_1 + X + \frac{X^2}{\omega_3^{\prime}-\omega _1^{\prime}} -\frac{X^2}{\omega_1^{\prime}-\omega _2^{\prime}}
+\frac{2\,X^3}{(\omega_3^{\prime}-\omega _1^{\prime})^2}
\notag\\
&\quad{}-\frac{3\,X^3}{(\omega_3^{\prime}-\omega _1^{\prime})\,(\omega_1^{\prime}-\omega _2^{\prime})}
+\frac{2\,X^3}{(\omega_1^{\prime}-\omega _2^{\prime})^2}
\notag\\
&\quad{}+\frac{10\, X^4}{3\,(\omega^\prime_3-\omega^\prime_1)^3}
-\frac{10\,X^4}{(\omega^\prime_3-\omega^\prime_1)^2\,(\omega^\prime_1-\omega^\prime_2)}
\notag\\
&\quad{}+\frac{10\,X^4}{(\omega^\prime_3-\omega^\prime_1)\,(\omega^\prime_1-\omega^\prime_2)^2}
-\frac{10\,X^4}{3\,(\omega^\prime_1-\omega^\prime_2)^3}. 
\label{app2}
\end{align}

\section{Numerical Evaluation}
\label{estimation}
\subsection{Estimation of Eigenfrequency and Higher-order Correction}
\label{estimation1}
In this section, we use the results of perturbative expansion to investigate the change in eigenfrequency caused by the effect of the one-way link graph and evaluate the effectiveness and properties of perturbative expansion. 
As a numerical example of equation \eqref{Omega}, we conduct numerical experiments using matrix $\bm{\Omega}(\epsilon)$, 
\begin{align}
  \bm{\Omega}(\epsilon) &= \bm{\Omega}_0+\epsilon \, \bm{\Omega}_\mathrm{I}\notag\\
  &=
  \begin{bmatrix}
    \omega_1&0&0\\
    0&\omega_2&0\\
    0&0&\omega_3
  \end{bmatrix}
  +
  \epsilon
  \begin{bmatrix}
  d_1&-a_1&0\\
  0&d_2&-a_2\\
  -a_3&0&d_3
  \end{bmatrix}
  \notag\\
  &=
  \begin{bmatrix}
    9&0&0\\
    0&6&0\\
    0&0&0
  \end{bmatrix}
  +
  \epsilon
  \begin{bmatrix}
  33/10&-5&0\\
  0&4/41&-4\\
  -4&0&16/15
  \end{bmatrix}. 
  \label{num_ex_m}
\end{align}
In the previous section, only the change in eigenfrequency $\hat{\omega}_1'$ due to perturbative expansion was shown. 
We can also consider the perturbative expansion for changes in the other eigenfrequencies $\hat{\omega}_2'$ and $\hat{\omega}_3'$. 
Following the $\hat{\omega}_1'$ analysis, we can examine $B_1$ and $C_1$ in \eqref{form} for the solution $\psi_1(t)$. 
However, there is a simpler way to obtain them as we describe next. 
In Figure \ref{mode}, changing the numbering of the states does not change their properties, so the result of the previous section with the cyclic replacement of subscripts $1 \rightarrow 3 \rightarrow 2 \rightarrow 1 \rightarrow \cdots$ and $X \rightarrow Y \rightarrow Z \rightarrow X \rightarrow \cdots$ is also valid. 
The replacement of the subscripts follows the correspondence in Table \ref{subscript}. 
The correspondence of $X, Y, Z$ obtained from Table \ref{subscript},  is shown in Table \ref{xyz}.

\begin{table}[tb]
\centering
\caption{Correspondence of subscripts.}
\vspace{2mm}
\begin{tabular}{c|c c c}
solution & \multicolumn{3}{c}{subscript}\\
\hline
$\psi_1(t)$ & $1$ & $2$ & $3$\\
\hline
$\psi_3(t)$ & $3$ & $1$ & $2$\\
\hline
$\psi_2(t)$ & $2$ & $3$ & $1$
\end{tabular}
\label{subscript}
\end{table}

\begin{table}[tb]
\caption{Correspondence of $X$, $Y$, $Z$.}
\vspace{2mm}
\centering
\begin{tabular}{c|c c c}
solution & \multicolumn{3}{c}{$X$, $Y$, $Z$}\\
\hline
$\psi_1(t)$ & $X$ & $Y$ & $Z$\\
\hline
$\psi_3(t)$ & $Y$ & $Z$ & $X$\\
\hline
$\psi_2(t)$ & $Z$ & $X$ & $Y$
\end{tabular}
\label{xyz}
\end{table}

By changing only the subscripts in expression \eqref{app0}, we obtain the lowest-order approximations, 
\begin{align}
    \hat{\omega}_3'\simeq\omega_3'+Y
    \label{app0_psi3}
\end{align}
\begin{align}
    \hat{\omega}_2'\simeq\omega_2'+Z. 
    \label{app0_psi2}
\end{align}

The higher-order corrections to the first eigenfrequency \eqref{app1} and \eqref{app2} can also be obtained for the other eigenfrequencies by replacing the subscripts according to Table \ref{subscript} and Table \ref{xyz}. 
By replacing the subscripts for \eqref{app1}, we obtain the corrections for the third eigenfrequency as follows: 
\begin{align}
\hat{\omega}_3 \simeq \omega^{\prime}_3 + Y + \frac{Y^2}{\omega_2^{\prime}-\omega _3^{\prime}} -\frac{Y^2}{\omega_3^{\prime}-\omega _1^{\prime}}. 
\label{app1_psi3}
\end{align}
Also, we obtain the corrections for the second eigenfrequency as follows: 
\begin{align}
\hat{\omega}_2 \simeq \omega^{\prime}_2 + Z + \frac{Z^2}{\omega_1^{\prime}-\omega _2^{\prime}} -\frac{Z^2}{\omega_2^{\prime}-\omega _3^{\prime}}
\label{app1_psi2}
\end{align}

Similarly, replacing the subscripts of the expression \eqref{app2}, we obtain the corrections for the third eigenfrequency as follows: 
\begin{align}
\hat{\omega}_3 &\simeq \omega^{\prime}_3 + Y + \frac{Y^2}{\omega_2^{\prime}-\omega _3^{\prime}} -\frac{Y^2}{\omega_3^{\prime}-\omega _1^{\prime}}
+\frac{2\,Y^3}{(\omega_2^{\prime}-\omega _3^{\prime})^2}
\notag\\
&\quad{}-\frac{3\,Y^3}{(\omega_2^{\prime}-\omega _3^{\prime})\,(\omega_3^{\prime}-\omega_1^{\prime})}
+\frac{2\,Y^3}{(\omega_3^{\prime}-\omega_1^{\prime})^2}
\notag\\
&\quad{}+\frac{10\, Y^4}{3\,(\omega^\prime_2-\omega^\prime_3)^3}
-\frac{10\,Y^4}{(\omega^\prime_2-\omega^\prime_3)^2\,(\omega^\prime_3-\omega^\prime_1)}
\notag\\
&\quad{}+\frac{10\,Y^4}{(\omega^\prime_2-\omega^\prime_3)\,(\omega^\prime_3-\omega^\prime_1)^2}
-\frac{10\,Y^4}{3\,(\omega^\prime_3-\omega^\prime_1)^3}.
\label{app2_psi3}
\end{align}
Also, we obtain the corrections for the second eigenfrequency as follows: 
\begin{align}
\hat{\omega}_2 &\simeq \omega^{\prime}_2 + Z + \frac{Z^2}{\omega_1^{\prime}-\omega _2^{\prime}} -\frac{Z^2}{\omega_2^{\prime}-\omega _3^{\prime}}
+\frac{2\,Z^3}{(\omega_1^{\prime}-\omega _2^{\prime})^2}
\notag\\
&\quad{}-\frac{3\,Z^3}{(\omega_1^{\prime}-\omega _2^{\prime})\,(\omega_3^{\prime}-\omega_1^{\prime})}
+\frac{2\,Z^3}{(\omega_3^{\prime}-\omega_1^{\prime})^2}
\notag\\
&\quad{}+\frac{10\, Z^4}{3\,(\omega^\prime_2-\omega^\prime_3)^3}
-\frac{10\,Z^4}{(\omega^\prime_2-\omega^\prime_3)^2\,(\omega^\prime_3-\omega^\prime_1)}
\notag\\
&\quad{}+\frac{10\,Z^4}{(\omega^\prime_2-\omega^\prime_3)\,(\omega^\prime_3-\omega^\prime_1)^2}
-\frac{10\,Z^4}{3\,(\omega^\prime_3-\omega^\prime_1)^3}. 
\label{app2_psi2}
\end{align}

\begin{table}[tb]
\centering
\caption{Parameters used in the numerical example of \eqref{num_ex_m}}
\vspace{2mm}
\begin{tabular}{c | c | l }
$\omega_\mu$ & $a_\mu$ & $\quad d_\mu$ \\
\hline
$\omega_1 = 9$ & $a_1 = 5$ & $d_1 = 33/10$ \\
$\omega_2 = 6$ & $a_2 = 4$ & $d_2 = 4/41$ \\
$\omega_3 = 0$ & $a_3 = 4$ & $d_3 = 16/15$ 
\end{tabular}
\label{parameter1}
\end{table}

\begin{figure}[tb]
  \centering
  \includegraphics[width=0.8\linewidth]{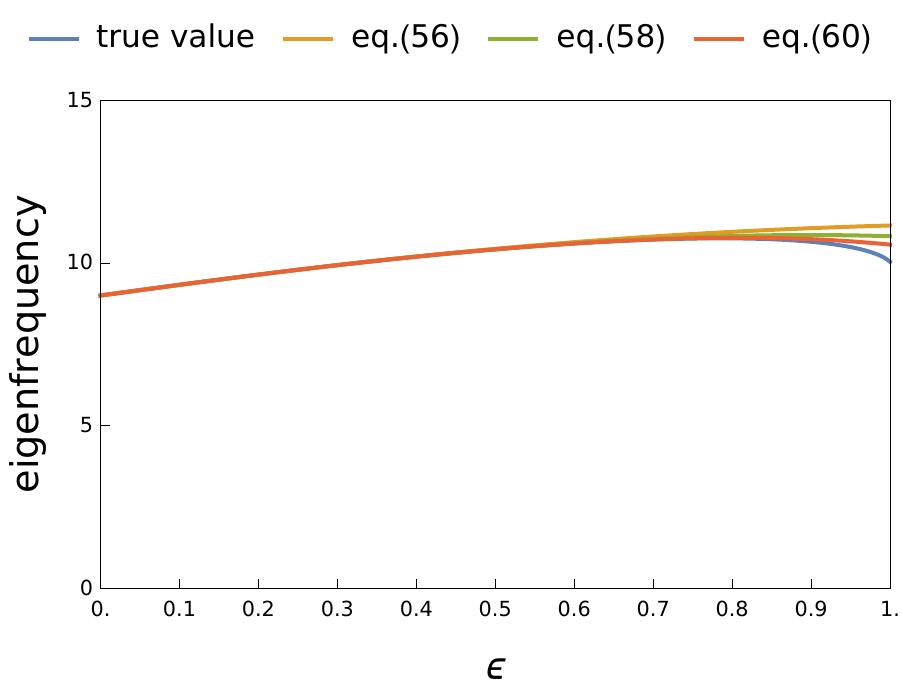}
  \caption{Estimated values of the first eigenfrequency}
  \label{w1}
\end{figure}

\begin{figure}[tb]
  \centering
  \includegraphics[width=0.8\linewidth]{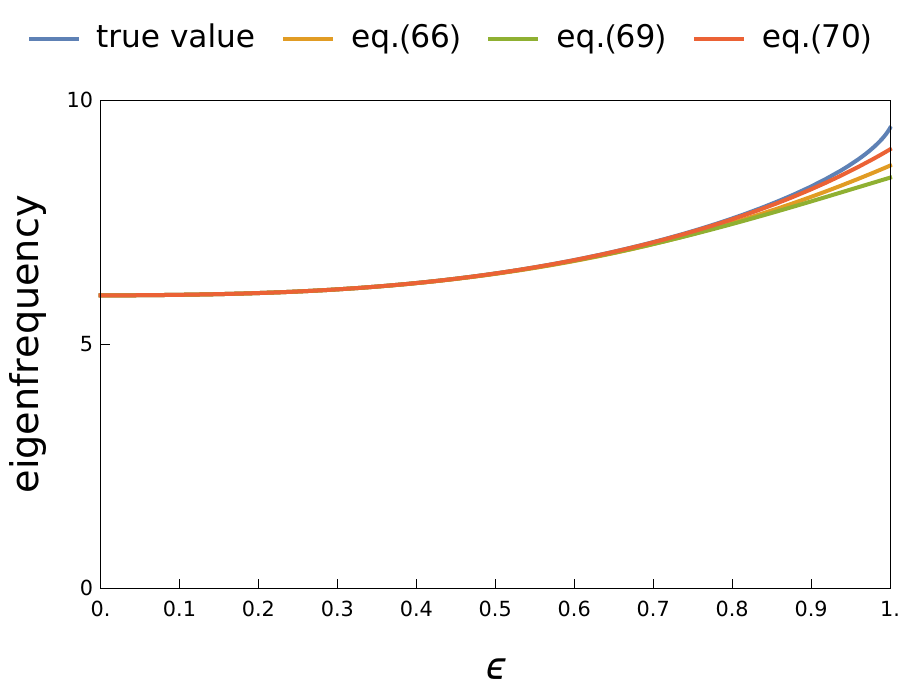}
  \caption{Estimated values of the second eigenfrequency}
  \label{w2}
\end{figure}

\begin{figure}[tb]
  \centering
  \includegraphics[width=0.8\linewidth]{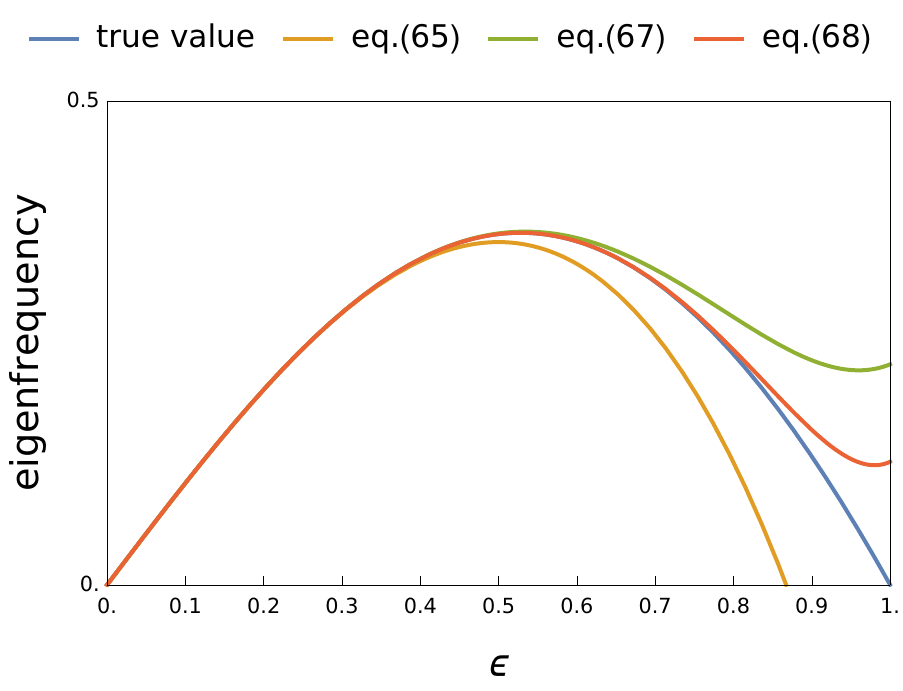}
  \caption{Estimated values of the third eigenfrequency}
  \label{w3}
\end{figure}

The estimated eigenfrequencies obtained by substituting the values determined from the expression \eqref{num_ex_m} into these expressions, with the value of parameter $\epsilon$ $(0 \le \epsilon \le 1)$, are shown in Figure ~\ref{w1}--\ref{w3}. 
The true eigenfrequency is plotted as the blue line for comparison. 
Each result shows a high estimation accuracy when the parameter $\epsilon$ is small, but the accuracy decreases as the parameter approaches $\epsilon\simeq 1$. 
However, the results with higher-order corrections show higher accuracy over a wider range of $\epsilon$.


\subsection{Accuracy of Eigenfrequency Estimation and Magnitude of Perturbation}
\label{estimation_higher}
The accuracy of the estimation of the eigenfrequency by the perturbative expansion improves when the absolute values of $X$, $Y$, and $Z$ at $\epsilon = 1$ are small. 
This is because if these absolute values are small, the perturbative expansion can be characterized  by just the terms of low order.

Without changing the $\bm{\Omega}_0$ in \eqref{num_ex_m}, consider two examples where $X$, $Y$, and $Z$ are small or large at $\epsilon = 1$; they are as follows: 
\begin{align}
  \bm{\Omega}(\epsilon) &= \bm{\Omega}_0+\epsilon \, \bm{\Omega}_\mathrm{I}\notag\\
  &=
  \begin{bmatrix}
    9&0&0\\
    0&6&0\\
    0&0&0
  \end{bmatrix}
  +
  \epsilon
  \begin{bmatrix}
  3/2&-1&0\\
  0&34/21&-2\\
  -1&0&1/40
  \end{bmatrix}, 
  \label{num_ex_s}
\end{align}
\begin{align}
  \bm{\Omega}(\epsilon) &= \bm{\Omega}_0+\epsilon \, \bm{\Omega}_\mathrm{I}\notag\\
  &=
  \begin{bmatrix}
    9&0&0\\
    0&6&0\\
    0&0&0
  \end{bmatrix}
  +
  \epsilon
  \begin{bmatrix}
  3&-6&0\\
  0&11/4&-7\\
  -5&0&2
  \end{bmatrix}. 
  \label{num_ex_l}
\end{align}
In these examples, the link weights $a_\mu$ $(\mu=1,\,2,\,3)$ are determined, and $d_\mu$ is determined to satisfy the condition that $\bm{\Omega}(1)$ has eigenvalue $0$ as a feature of the Laplacian matrix. 
Table~\ref{table:XYZ} shows the values of $|X|$, $|Y|$, and $|Z|$ at $\epsilon = 1$ for models \eqref{num_ex_m}, \eqref{num_ex_s}, and \eqref{num_ex_l}. 
The model of \eqref{num_ex_l} is strongly influenced by $\bm{\Omega}_\mathrm{I}$, and thus non-real eigenfrequencies appear at $\epsilon=1$. 
In other words, when $\epsilon=0$, all eigenfrequencies are real, but when $\epsilon$ exceeds a certain value, non-real eigenfrequencies appear.

\begin{table}[tb]
\centering
\caption{Parameters given in a numerical example of  \eqref{num_ex_s}}
\vspace{2mm}
\begin{tabular}{c | c | l }
$\omega_\mu$ & $a_\mu$ & $\quad d_\mu$ \\
\hline
$\omega_1 = 9$ & $a_1 =1$ & $d_1 = 3/2$ \\
$\omega_2 = 6$ & $a_2 = 2$ & $d_2 = 34/21$ \\
$\omega_3 = 0$ & $a_3 = 1$ & $d_3 = 1/49$ 
\end{tabular}
\label{parameter2}
\end{table}

\begin{table}[tb]
\centering
\caption{Parameters used in the numerical example of \eqref{num_ex_l}}
\vspace{2mm}
\begin{tabular}{c | c | l }
$\omega_\mu$ & $a_\mu$ & $\quad d_\mu$ \\
\hline
$\omega_1 = 9$ & $a_1 = 6$ & $d_1 = 3$ \\
$\omega_2 = 6$ & $a_2 = 7$ & $d_2 = 11/4$ \\
$\omega_3 = 0$ & $a_3 = 5$ & $d_3 = 2$ 
\end{tabular}
\label{parameter3}
\end{table}

\begin{table}[tb]
\centering
\caption{Values of $|X|$, $|Y|$, and $|Z|$ of each model for $\epsilon = 1$}
\vspace{2mm}
\begin{tabular}{c||c|c|c}
model & $|X|$ & $|Y| $ & $|Z| $ \\
\hline
\eqref{num_ex_m} for $\epsilon = 1$ & 1.148 & 1.416 & 2.564 \\
\eqref{num_ex_s} for $\epsilon = 1$ & 0.066 & 0.025 & 0.091 \\
\eqref{num_ex_l} for $\epsilon = 1$ & 6.462 & 3.111 & 9.573 
\end{tabular}
\label{table:XYZ}
\end{table}

\begin{figure}[tb]
  \centering
  \includegraphics[width=0.8\linewidth]{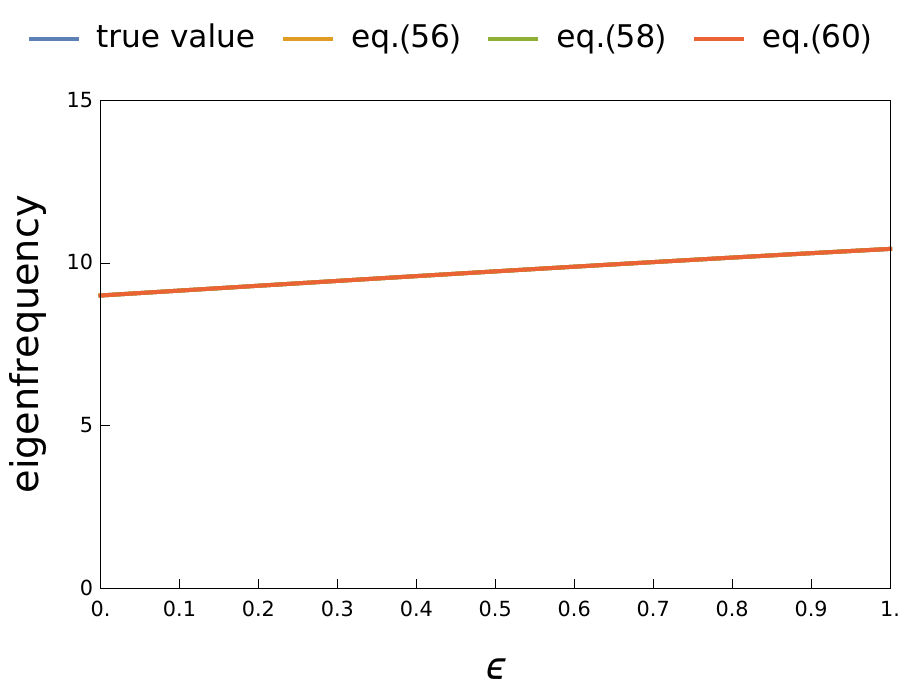}
  \caption{Comparison of estimation accuracy with higher-order corrections for the first eigenfrequency of the numerical example of \eqref{num_ex_s}}
  \label{w1_high_s}
\end{figure}

\begin{figure}[tb]
  \centering
  \includegraphics[width=0.8\linewidth]{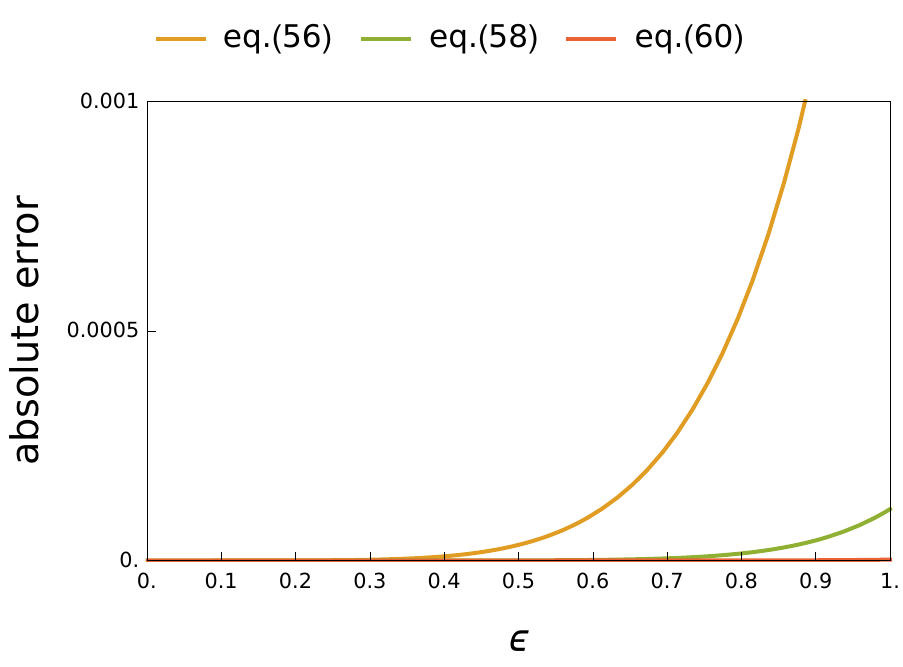}
  \caption{Comparison of absolute error of estimation values for the first eigenfrequency of the numerical example of  \eqref{num_ex_s}}
  \label{w1_high_abs_err_s}
\end{figure}

\begin{figure}[tb]
  \centering
  \includegraphics[width=0.8\linewidth]{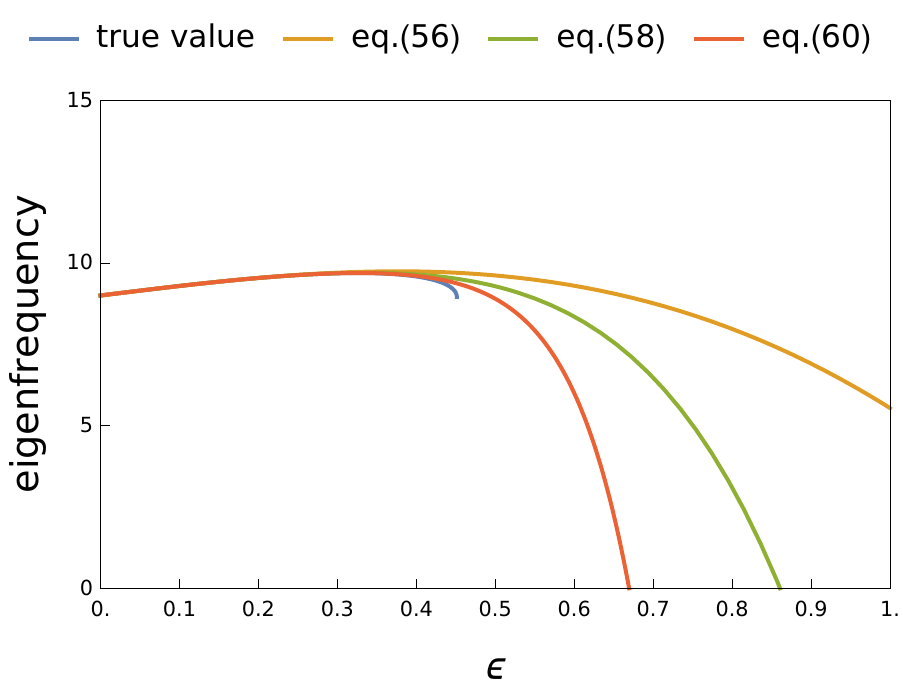}
  \caption{Comparison of estimation accuracy with higher-order corrections for the first eigenfrequency of the numerical example of \eqref{num_ex_l}}
  \label{w1_high_l}
\end{figure}

Figure \ref{w1_high_s} shows the estimation results of the first eigenfrequency for the numerical example \eqref{num_ex_s}.
The true value is shown in blue, and the estimation results using perturbations \eqref{app0}, \eqref{app1}, and \eqref{app2} are shown in ascending order.
The lines virtually overlap, and compared to Figure \ref{w1}, Figure \ref{w1_high_s} provides high estimation accuracy even at $\epsilon=1$, indicating that the perturbative calculation is working effectively.

Figure \ref{w1_high_abs_err_s} shows a plot of the error between the true value and each approximate calculation for a precise evaluation of the estimation accuracy of the perturbative calculation.
This result shows that the error decreases as the order of the perturbative calculation increases, and the estimation result by \eqref{app2} has extremely high estimation accuracy even for $\epsilon=1$.

On the other hand, Fig.~\ref{w1_high_l} shows a similar estimation result for numerical example \eqref{num_ex_l}. 
Since the true value only exists as a real value in the region where $\epsilon$ is approximately $\epsilon<0.45$, the results of the perturbative calculation for $\epsilon$ larger than that are meaningless.

\begin{figure}[tb]
  \centering
  \includegraphics[width=0.8\linewidth]{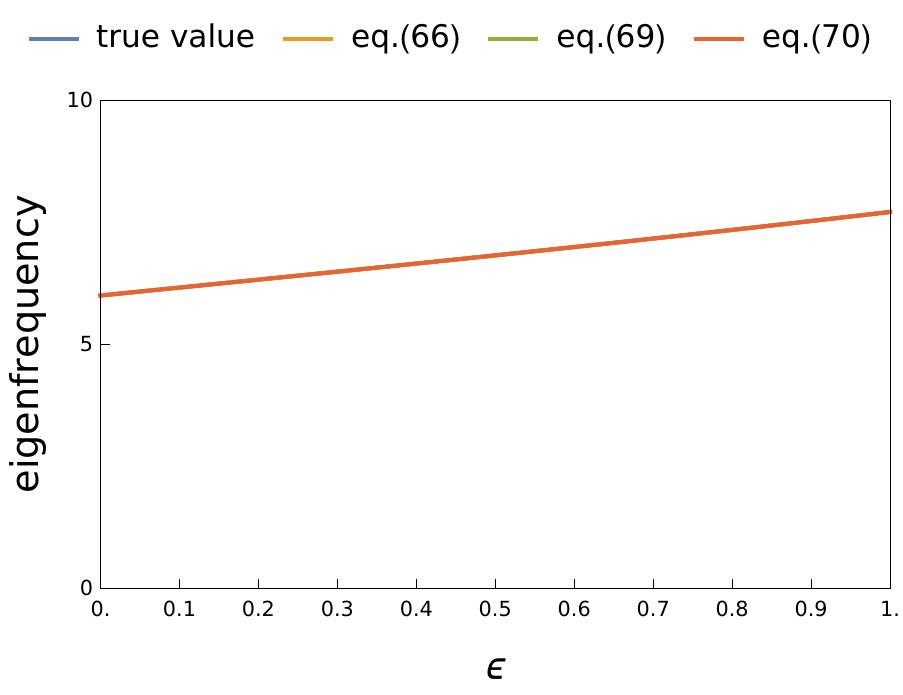}
  \caption{Comparison of estimation accuracy with higher-order corrections for the second eigenfrequency of numerical example \eqref{num_ex_s}}
  \label{w2_high_s}
\end{figure}

\begin{figure}[tb]
  \centering
  \includegraphics[width=0.8\linewidth]{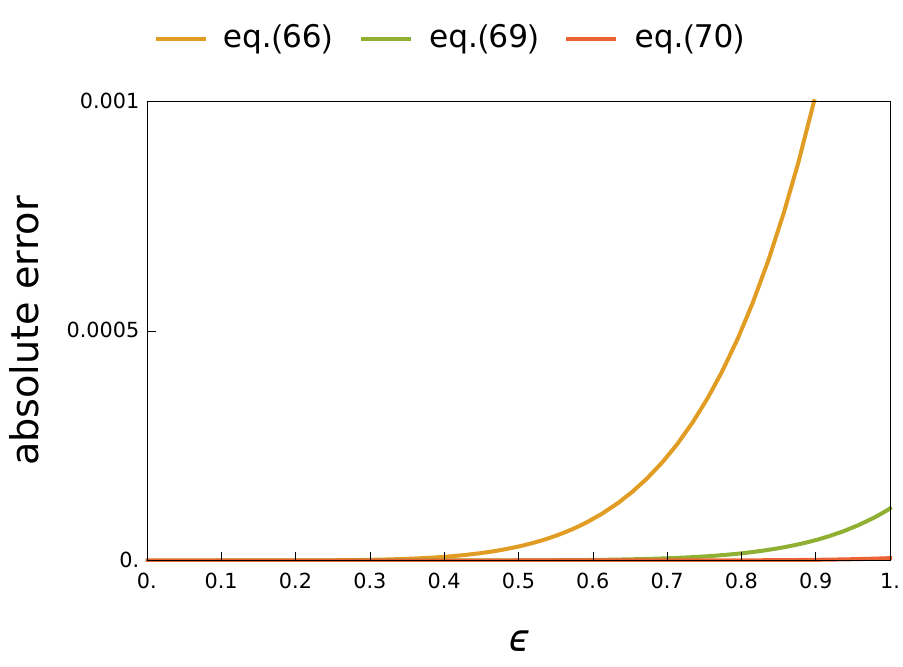}
  \caption{Comparison of absolute error of estimation values for the second eigenfrequency of numerical example \eqref{num_ex_s}}
  \label{w2_high_abs_err_s}
\end{figure}

Similar evaluations were performed for the second and third eigenfrequencies. 
Figure \ref{w2_high_s} shows the estimation results of the second eigenfrequency for numerical example \eqref{num_ex_s}.
The true value is shown in blue, and the estimation results by the perturbative expansions \eqref{app0}, \eqref{app1}, and \eqref{app2} are shown in ascending order. 
As in the case of the first eigenfrequency, the lines virtually overlap and high estimation accuracy is achieved even for $\epsilon=1$. 
The error between the true value and each approximation is shown in Figure \ref{w2_high_abs_err_s}. It  shows that the estimation result yielded by \eqref{app2} is extremely accurate even for $\epsilon=1$. 
The results of the perturbative calculation for approximately $\epsilon > 0.45$ are meaningless and the non-real eigenfrequencies cannot be estimated.

\begin{figure}[tb]
  \centering
  \includegraphics[width=0.8\linewidth]{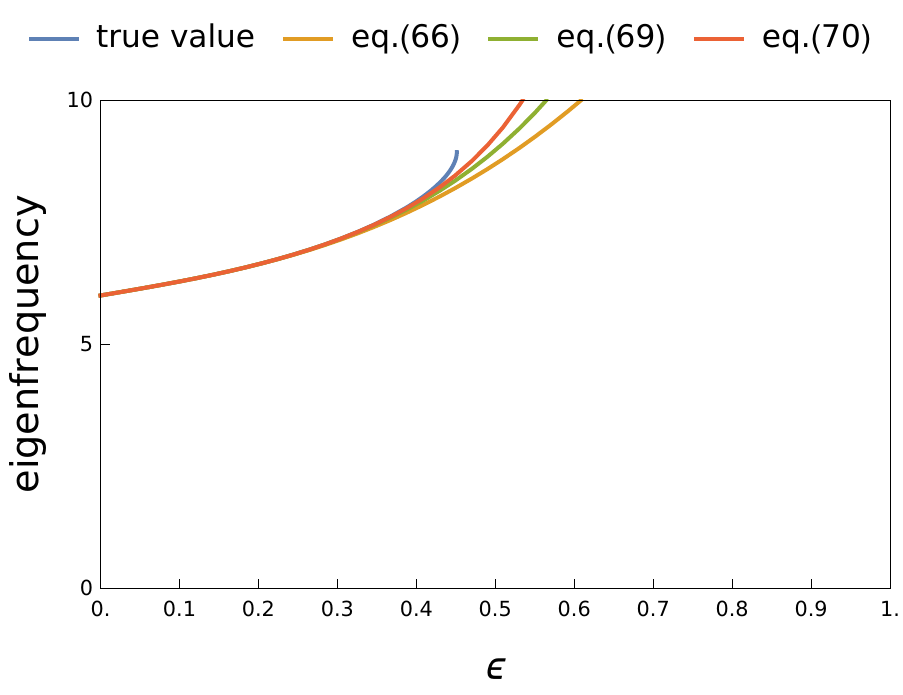}
  \caption{Comparison of estimation accuracy with higher-order corrections for the second eigenfrequency for numerical example \eqref{num_ex_s}}
  \label{w2_high_l}
\end{figure}

\begin{figure}[tb]
  \centering
  \includegraphics[width=0.8\linewidth]{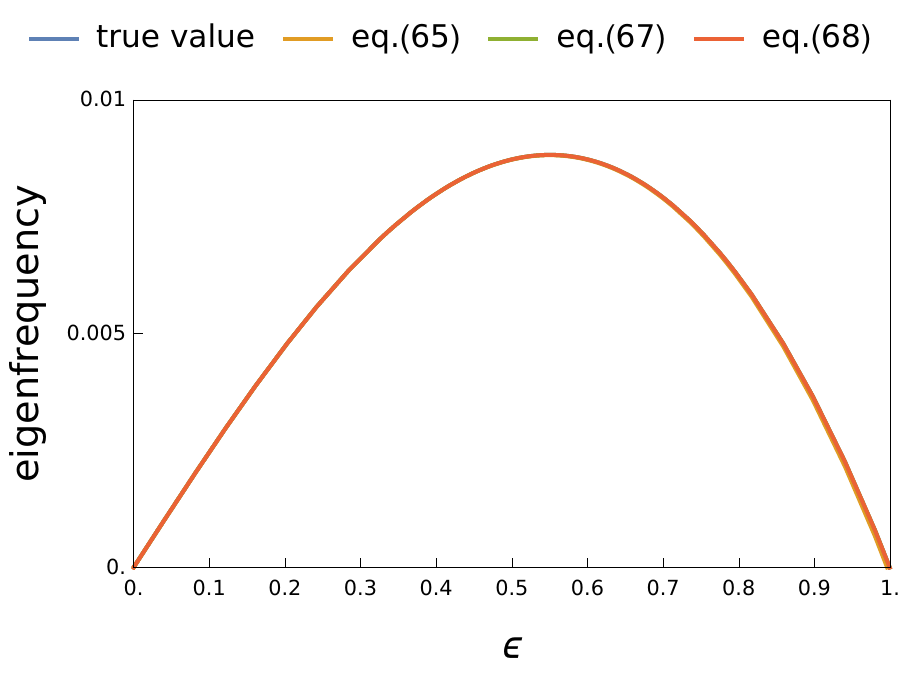}
  \caption{Comparison of estimation accuracy with higher-order corrections for the third eigenfrequency of numerical example \eqref{num_ex_s}}
  \label{w3_high_s}
\end{figure}

\begin{figure}[tb]
  \centering
  \includegraphics[width=0.8\linewidth]{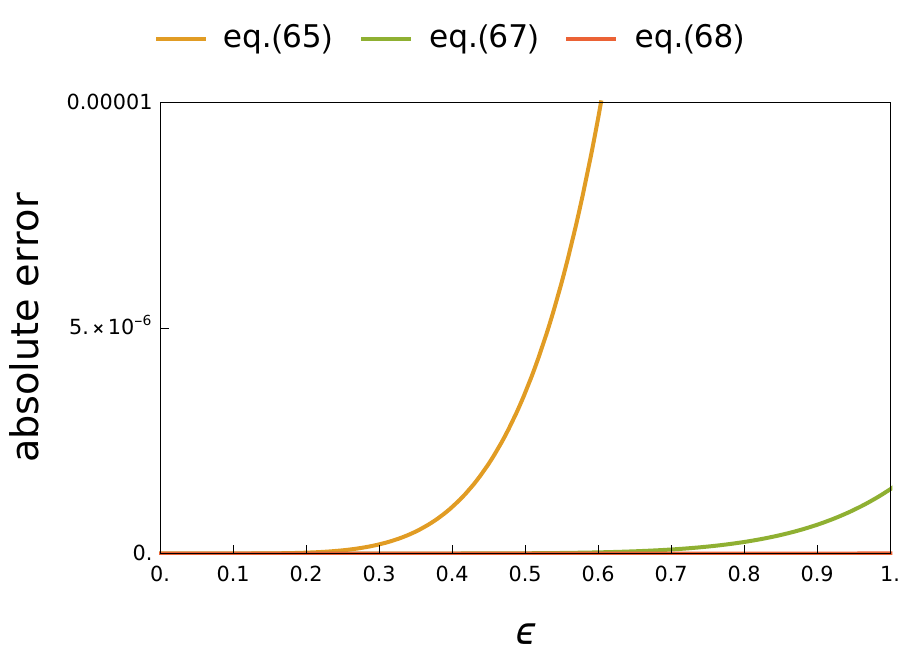}
  \caption{Comparison of absolute error of estimation values for the third eigenfrequency of numerical example \eqref{num_ex_s}}
  \label{w3_high_abs_err_s}
\end{figure}

\begin{figure}[tb]
  \centering
  \includegraphics[width=0.8\linewidth]{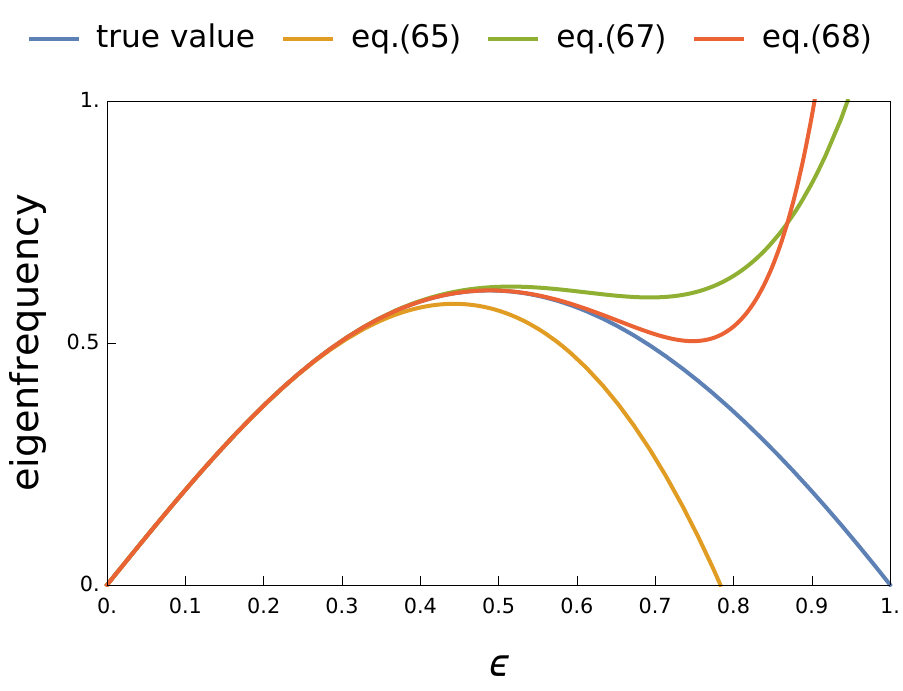}
  \caption{Comparison of estimation accuracy with higher-order corrections for the third eigenfrequency of numerical example \eqref{num_ex_l}}
  \label{w3_high_l}
\end{figure}

Figure \ref{w3_high_s} shows the estimation results of the third eigen frequency for numerical example \eqref{num_ex_s}. 
The true value is shown in blue, and the estimation results by perturbative expansion \eqref{app0}, \eqref{app1}, and \eqref{app2} are shown in ascending order. 
Similar to the previous results, the lines virtually overlap, and high estimation accuracy is achieved even for $\epsilon=1$. 
The error between the true value and each approximation is shown in Figure \ref{w3_high_abs_err_s}. It shows that the estimation result of \eqref{app2} is extremely accurate even for $\epsilon=1$. 
Figure \ref{w3_high_l} shows the same estimation result for numerical example \eqref{num_ex_l}. 
Since the third eigenfrequency is $0$ (i.e., a real value), the results of the perturbative calculation are meaningful even for $\epsilon = 1$, but the estimation accuracy is not good. 
The reason for this is that the higher-order terms in the perturbative expansion have a large effect on this numerical example, and the approximation with finite order does not work effectively.

The above evaluation experiments confirm that the accuracy of eigenfrequency estimation is improved by higher-order corrections and that the perturbative expansion at finite order works effectively when the influence of $\bm{\Omega}_\mathrm{I}$ is small. Moreover, the change in eigenfrequency can be estimated very accurately.

\section{Conclusion}
This paper evaluated the use of perturbative expansion for analyzing solutions of the fundamental equation of the oscillation model; the goal was to develop a simple mode-coupled model up to infinite order. 
By using hypergeometric series, we succeeded in systematically expressing the perturbative expansion of infinite order. 
Moreover, by using an exponential function to express the perturbative expansion, we showed how to estimate, in perturbation theory, the change in eigenfrequency.

Experiments on the estimation of the eigenfrequencies using perturbation theory showed that the eigenfrequencies can be estimated with high accuracy for networks where the influence of $\bm{\Omega}_\mathrm{I}$ is small. 

The perturbation theoretic treatment of user dynamics on networks shown in this paper is important for understanding the dynamics occurring in networks based on causal relationships.

\section*{Acknowledgement}
This research was supported by Grant-in-Aid for Scientific Research 19H04096, 20H04179, and 21H03432 from the Japan Society for the Promotion of Science (JSPS), and TMU local 5G research support.


\begin{thebibliography}{99}

\bibitem{MasakiAIDA20182017EBN0001}
M.~Aida, C.~Takano, and M.~Murata, 
``Oscillation model for describing network dynamics caused by asymmetric node interaction,''  {\it IEICE Transactions on Communications}, 
vol.~E101-B, no.~1, pp.~123--136, January 2018.

\bibitem{freeman1978centrality}
L.C.~Freeman, 
``Centrality in social networks conceptual clarification,'' 
{\it Social Networks}, vol.~1, no.~3, pp.~215--239, 1978.

\bibitem{borgatti2009network}
S.P.~Borgatti, A.~Mehra, D.J.~Brass, and G.~Labianca.
``Network analysis in the social sciences,'' 
{\it Science}, vol.~323, no.~5916, pp.~892--895, 2009.

\bibitem{TakanoIEICE.T.2018}
C.~Takano and M.~Aida, 
``Revealing of the underlying mechanism of different node centralities based on oscillation dynamics on networks,'' 
{\it IEICE Transactions on Communications},  vol.~E101.B, no.~8, pp.~1820--1832, 2018.

\bibitem{aidaASONAM2017}
M.~Aida, C.~Takano, and M.~Murata.
``Dynamical model of flaming phenomena in on-line social networks,'' 
{\it the 2017 IEEE/ACM International Conference on Advances in Social Networks Analysis and Mining 2017 (ASONAM '17)}, pp.~1164--1171, 2017. 

\bibitem{aida2018generation}
M.~Aida, C.~Takano, and M.~Murata, 
``Generation mechanism of flaming phenomena in on-line social networks described by perturbation of asymmetric link effects,'' 
{\it 2018 IEEE/IFIP Network Operations and Management Symposium (NOMS 2018)}, pp.~1--4, 2018

\bibitem{pastor2001epidemic}
R.~Pastor-Satorras, A.~Vespignani,
``Epidemic spreading in scale-free networks,"
{\it Physical review letters},
vol.~86, no.~14, pp.~3200--3203, 2001.

\bibitem{newman2002spread}
M.E.J.~Newman,
``Spread of epidemic disease on networks,"
{\it Physical review E}
vol.~66, no.~1, 2002, Art.no.~016128.

\bibitem{nekovee2007theory}
M.~Nekovee, Y.~Moreno, G.~Bianconi, M.M.~Ginestra,
``Theory of rumour spreading in complex social networks,"
{\it Physica A: Statistical Mechanics and its Applications}
vol.~374, no.~1, pp.~457--470, 2007.

\bibitem{cannarella2014epidemiological}
J.~Cannarella, J.A.~Spechler,
``Epidemiological modeling of online social network dynamics,"
{\it arXiv preprint arXiv:1401.4208}
2014.

\bibitem{olfati2007consensus}
R.~Olfati-Saber, J.A.~Fax, R.M.~Murray,
``Consensus and cooperation in networked multi-agent systems,"
{\it Proceedings of the IEEE},
vol.~95, no.~1, pp.~215--233, 2007.

\bibitem{ren2005consensus}
W.~Ren, R.W.~Beard,
``Consensus seeking in multiagent systems under dynamically changing interaction topologies,"
{\it IEEE Transactions on automatic control}
vol.~50, no.~5, pp.~655--661, 2005.

\newpage

\bibitem{baumann2020modeling}
F.~Baumann, P.~Lorenz-Spreen, I.M.~Sokolov, M.~Starnini,
``Modeling echo chambers and polarization dynamics in social networks,"
{\it Physical Review Letters}
vol.~124, no.~4, 2020, Art.no.048301.

\bibitem{tornberg2018echo}
P.~T{\"o}rnberg,
``Echo chambers and viral misinformation: Modeling fake news as complex contagion,"
{\it PloS one}
vol.~13, no.~9, 2018, Art.no.~e0203958.

\bibitem{iacopini2018network}
I.~Iacopini, S.~Milojevi{\'c}, V.~Latora,
``Network dynamics of innovation processes,"
{\it Physical review letters}
vol.~120, no.~4, 2018, Art.no.~048301.

\bibitem{gleeson2014simple}
J.P.~Gleeson, D.~Cellai, J.P.~Onnela, M.A.~Porter, F.~Reed-Tsochas,
``A simple generative model of collective online behavior,"
{\it Proceedings of the National Academy of Sciences}
vol.~111, no.~29, pp.~10411--10415, 2014.

\bibitem{lerman2010information}
K.~Lerman, R.~Ghosh,
``Information contagion: An empirical study of the spread of news on digg and twitter social networks,"
{\it Proceedings of the International AAAI Conference on Web and Social Media}
vol.~4, no.~1, 2010.

\bibitem{bakshy2011everyone}
E.~Bakshy, J.M.~Hofman, W.A.~Mason, D.J~Watts,
``Everyone's an influencer: quantifying influence on twitter,"
{\it Proceedings of the fourth ACM international conference on Web search and data mining}
pp.~65--74, 2011.

\bibitem{doerr2012rumors}
B.~Doerr, M.~Fouz, and T.~Friedrich, 
``Why rumors spread so quickly in social networks,'' 
{\it Communications of the ACM}, vol.~55, no.~6, pp.~70--75, 2012.

\bibitem{bakshy2012role}
E.~Bakshy, I.~Rosenn, C.~Marlow, and L.~Adamic, 
``The role of social networks in information diffusion,'' 
{\it The 21st International Conference on World Wide Web}, pp.~519--528, 2012

\bibitem{myers2014bursty}
S.A.~Myers and J.~Leskovec, 
``The bursty dynamics of the twitter information network,'' 
{\it The 23rd International Conference on World Wide Web}, pp.~913--924, 2014.

\bibitem{centola2010spread}
D.~Centola, 
``The spread of behavior in an online social network experiment,'' 
{\it Science}, vol.~329, no.~5996, pp.~1194--1197, 2010.

\bibitem{aida2020book}
M.~Aida, 
{\it Introduction to Network Dynamics}, Morikita Publishing Co.~Ltd., 2020. (in Japanese)

\bibitem{kronenburg2011binomial}
M.J.~Kronenburg, 
``The binomial coefficient for negative arguments,'' 
{\it arXiv:1105.3689}, 2011.

\end{thebibliography}
\end{document}